\documentstyle[11pt,mmn,amsmath,amssymb,amsfonts,theorem,float]{article}

\newtheorem{theorem}{Theorem}[section]

\newtheorem{cor}[theorem]{Corollary}
\newtheorem{lemma}[theorem]{Lemma}
\newtheorem{algorithm}[theorem]{Algorithm}

\newtheorem{example}[theorem]{Example}

\def\startpage{1}       
\def\runauthor{X.S. Gao, Z.Y. Huang}

\def\runtitle{Characteristic Set Method in Finite Fields}

\newenvironment{Proof}{\noindent {\em Proof:\ }}{$\square$}

\newcommand{\SPC}{\hspace*{15pt}}

\def\bref#1{(\ref{#1})}

\def\F{{\mathbb{F}}}

\def\Q{{\mathbb{Q}}}
\def\C{{\mathbb{C}}}
\def\R{{\mathbb{R}}}

\def\X{{\mathbb{X}}}
\def\Y{{\mathbb{Y}}}
\def\U{{\mathbb{U}}}

\def\PS{{\mathbb{P}}}
\def\QS{{\mathbb{Q}}}
\def\HS{{\mathbb{H}}}

\def\sat{\hbox{\rm{sat}}}
\def\Zero{\hbox{\rm{Zero}}}
\def\zero{\hbox{\rm{Zero}}}
\def\hzero{\hbox{\rm{Zero}}_{q}}
\def\tzero{\hbox{\rm{Zero}}_{2}}

\def\class{\hbox{\rm{cls}}}
\def\cls{\hbox{\rm{cls}}}
\def\dim{\hbox{\rm{dim}}}
\def\lvar{\hbox{\rm{lvar}}}
\def\initial{\hbox{\rm{init}}}

\def\res{\hbox{\rm{resl}}}
\def\prem{\hbox{\rm{prem}}}

\def\deg{\hbox{\rm{deg}}}
\def\tdeg{\hbox{\rm{tdeg}}}
\def\term{\hbox{\rm{term}}}

\def\OR{{\mathcal{R}}}
\def\OJ{{\mathcal{J}}}
\def\OO{{\mathcal{O}}}
\def\OS{{\mathcal{S}}}
\def\OI{{\mathcal{I}}}

\def\ID{\hbox{\rm{I}}}

\def\I{{{\bf{I}}}}
\def\A{{{\mathcal{A}}}}
\def\B{{{\mathcal{B}}}}
\def\C{{{\mathcal{C}}}}

\def\AS{{{\mathcal{A}}}}

\def\res{\hbox{\rm{resl}}}
\def\resl{\hbox{\rm{resl}}}

\def\BI{{\mathbf{I}}}

\def\PS{{\mathbb{P}}}
\def\SS{{\mathbb{S}}}

\begin{document}
\title{{\large\bf Characteristic Set Algorithms for  Equation Solving in Finite Fields \\and Applications in Cryptanalysis$^1$}
 }
\author{Xiao-Shan Gao and Zhenyu Huang\\
Key Laboratory of Mathematics Mechanization\\
Institute of Systems Science, AMSS, Chinese Academy of Sciences}

\maketitle

\footnotetext[1]{Partially supported by a National Key Basic
Research Project of China and a grant from NSFC.}

\begin{abstract}
Efficient characteristic set methods for computing solutions of
polynomial equation systems in a finite field are proposed.
The concept of proper triangular sets is introduced  and an explicit formula
for the number of solutions of a proper and monic (or regular) triangular set is given.
An improved zero decomposition algorithm which can be used to reduce the
zero set of an equation system in general form to the union of zero
sets of monic proper triangular sets is proposed.
As a consequence, we can give an explicit formula for the number of
solutions of an equation system.
Bitsize complexity for the algorithm is given in the case of Boolean polynomials.
We also give a multiplication free characteristic set method for Boolean polynomials,
where the sizes of the polynomials are effectively controlled.
The algorithms are implemented in the case of Boolean polynomials
and extensive experiments show that they are quite efficient for
solving certain classes of Boolean equations.

\end{abstract}

\section{Introduction}

Solving polynomial equations  in finite fields plays a fundamental
role in many important fields such as coding theory, cryptology,
 and analysis of computer hardware. To find efficient
algorithms to solve such equations is a central issue both in
mathematics and in computer science (see Problem 3 in \cite{smale}
and Section 8 of \cite{ecrypt}).
Efficient algebraic algorithms for solving equations in finite
fields have been developed, such as the Gr\"obner basis methods
\cite{bardet1,polybori,f4,f5,fa1,kapur0,vg1,sato} and the XL
algorithm and its improved versions \cite{xl1}.

The {\bf characteristic set (CS)} method is a tool for studying
polynomial, algebraic differential, and algebraic difference
equation systems
\cite{la1,boul,rody1,chou1,cade10,lift,gallo,gluo,hub,kalk1,kapur1,la2,lindd1,maza1,moller,wang,wu-basic,yz1}.
The idea of the method is reducing equation systems in general form
to equation systems in the form of triangular sets.
%
With this method, solving an equation system can be reduced to
solving univariate equations in cascaded form. In the case of finite
fields, univariate equations can be solved with Berlekamp's
algorithm \cite{hac}.
%
The CS method can also be used to compute the dimension, the degree,
and the order for an equation system, to solve the radical ideal
membership problem, and to prove theorems from elementary and
differential geometries \cite{wu-ebook}.

In most existing work on CS methods, the zeros of the equations are
taken in an algebraically closed field which is infinite. These
methods can also be used to solve equations in finite fields. But,
they do not take into the account of the special properties of the
finite fields and thus are not efficient for solving equations in
finite fields.
%
%
In this paper, we propose efficient CS methods to solve
equations in the general finite field $\F_q$ with $q$ elements. More
precisely, we will develop efficient CS algorithms for polynomial
systems in the ring
$$\R_q= \F_q[x_1,\ldots,x_n]/(\HS)$$
where $\HS = \{x_1^q-x_1,\ldots,x_n^q-x_n\}$.
Due to the special property of $\R_q$, the proposed CS methods are
more efficient and have better properties  than the general CS
method.

A triangular set  may have no solutions in a finite field. For
instance, $x^2+1=0$ has no solution in the finite field $\F_3$.
To avoid this problem, we introduce the concept of proper triangular
sets and prove that proper triangular sets are square-free. We also
give an explicit formula for the number of solutions of a monic and proper
triangular set.
We modify the definition of regular triangular sets \cite{la1,rody1,yz1}
in $\R_q$ and give an exact upper bound for the number of solutions
of a regular and proper triangular set.

We propose an improved zero decomposition algorithm which allows us
to decompose the zero set of a polynomial equation system in $\R_q$
as the disjoint union of the zero sets of proper and monic triangular sets. As
a consequence, we can give an explicit formula for the number of
solutions of the equation system.
%
We prove that our elimination procedure to compute a triangular set needs a
polynomial number of polynomial multiplications, which is not valid for the general CS method.

An element in $\R_2$ is called a Boolean polynomial. Solving Boolean
polynomial systems is especially important and more methods are
available.
This paper will focus on CS methods.
We show that for Boolean polynomial equations, the CS method
proposed in this paper and that proposed in \cite{gaof2} for Boolean
polynomials could be further improved.
First, we give a bitsize complexity for the zero decomposition
algorithm proposed in this paper. This is the first complexity
analysis for the zero decomposition algorithm. The results in
\cite{gallo} are only for the procedure to compute one CS, which is
called the well-ordering procedure by Wu \cite{wu-basic}.

We also present a multiplication-free CS algorithm in $\R_2$, where
the size of the polynomials occurring in the well-ordering procedure
is bounded by the size of the input polynomial system and the worst
case bitsize complexity of the algorithm is roughly $O(n^d)$, where $n$
is the number of indeterminates and $d$ the degree of the input polynomials.
This result is surprising, because repeated additions of polynomials
can also generate polynomials of exponential sizes. In the general
CS method, the size of the polynomials is exponential \cite{gallo}.
Our result also means that for a small $d$, the well-ordering
procedure is a polynomial-time algorithm in $n$.
The bottle neck problem of intermediate expression swell is
effectively avoided for certain classes of problems due to the low
complexity of the well-ordering procedure and the usage of SZDD
\cite{zdd}. Our experimental results also support this observation.


The algorithms are implemented in the case of Boolean polynomials.
We conduct extensive experiments of our methods for three kinds of
polynomial systems. These systems are generated in totally different
ways, but they all have the block triangular structure. By block
triangular structure, we mean that the polynomial set can be divided
into disjoint sets such that each set consists of polynomials with
the same leading variable and different sets have different leading
variables. Polynomial sets generated in many classes of stream
ciphers are in triangular block form.
The experiments show that our improved algorithm is very effective
for solving these polynomial equations comparing to existing
methods.
%
%
We do not claim that our algorithm is faster in all cases. For
instance, the first HFE Challenge, which was solved by the Gr\"obner
basis algorithm \cite{hfe-fg,hfe-patarin}, can not be solved by our
algorithm.


The rest of this paper is organized as follows. In Section 2, we
introduce the notations. In Section 3, we prove properties for the
proper triangular sets. In Section 4, we present the improved zero
decomposition algorithm.
In Section 5, we present a CS algorithm in $\R_2$. In Section 6, we
present the experimental results. In Section 7, conclusions are
presented.

\section{Notations and Preliminary Results}

Let $p$ be a prime number and $q=p^k$ for a positive integer $k$.
$\F_q$ denotes the finite field with $q$ elements. For an algebraic
equation, we will consider the problem of finding its solutions in
$\F_q$. Let $\X=\{x_1,\ldots,x_n\}$ be a set of indeterminates.
Since we only consider solutions in $\F_q$, we can work in the ring
$$\R_q=\F_q[\X]/(\HS)$$
where
\begin{equation} \label{H}
\HS=\{x_1^q-x_1,x_2^q-x_2,\ldots,x_n^q-x_n\}.
\end{equation}
When we want to emphasize the variables, we use the notation
$\R_q[x_1,\ldots,x_n]$ instead of $\R_q$. It is easy to see that
$\R_q$ is not an integral domain. For any $\alpha\in \F_q$,
$x_i-\alpha$ is a zero divisor in $\R_q$. An element $P$ in $\R_q$
has the following {canonical representation}:

\begin{equation}\label{eq-1}
P = \alpha_sM_s + \cdots + \alpha_0M_0, \quad \alpha_i\in \F_q,
\end{equation}
where $M_i$ is a monomial and deg$(M_i,x_j)\leq q-1$ for any
$j$\,.
%
%
We still call an element in $\R_q$ a {\bf  polynomial}.
In this paper, a polynomial is always in its canonical
representation.

Let $\PS$ be a set of polynomials  in $\R_q$. We use $\hzero(\PS)$
to denote the common zeros of the polynomials  in $\PS$ in the
affine space $\F_q^n$, that is,
$$\hzero(\PS)= \{(a_1,\ldots,a_n), a_i\in \F_q, s.t., \forall P\in \PS, P(a_1,\ldots,a_n)=0  \}.$$
In this paper, when we say a {\bf  variety} in $\F_q^n$, we mean
$\hzero(\PS)$ for some $\PS \subseteq \R_q[x_1,\ldots,x_n]$. Let $D$
be a polynomial in $\R_q$. We define a {\bf  quasi variety} to be
$$\hzero(\PS/D)= \hzero(\PS)\setminus \hzero(D).$$

Let $\PS$ be a set of polynomials  in $\F_q[\X]$. Denote the zeros
of $\PS$ in an algebraically closed extension of $\F_q$ as
$\Zero(\PS)$.
We use $\overline{\PS}$ to denote the image of $\PS$ under the
natural ring homomorphism:
$$\F_q[\X] \Rightarrow \R_q.$$
We will give some preliminary results about the polynomials  in
$\R_q$.

\begin{lemma}
\label{lm-1} Use the notations just introduced. We have
$\Zero(\PS\cup\HS) = \hzero(\overline{\PS})$, where $\HS$ is
defined in \bref{H}.
\end{lemma}
\begin{Proof}
Let $P\in \PS$. By the definition, we have  $P = \overline{P} +
\sum_i B_i(x_i^q - x_i)$, where $B_i$ are some polynomials. Note
that any zero in  $\hzero(\overline{\PS})$ is also a zero of
$x_i^q-x_i$. Then the formula to be proved is a direct consequence
of the above relation between $P$ and $\overline{P}$.
\end{Proof}

%
%

\begin{lemma}\label{lm-2}
Let P be a polynomial in $\R_q$. We have $P^q=P$.
\end{lemma}
\begin{Proof}
Since $x^q_i=x_i$, for any monomial $m$ in $\R_q$ we have $m^{q}=m$.
Let $P = \sum_i \alpha_im_i$ where $m_i$ are monomials and
$\alpha_i\in \F_q$. Then $P^q=(\sum_i \alpha_i m_i)^{q}=\sum_i
\alpha^q_im_i^q = \sum_i\alpha_i m_i =P$.
\end{Proof}

\begin{lemma}\label{lm-3}
Let $\ID$ be a polynomial ideal in $\R_q$. Then $\ID$ is a radical
ideal.
\end{lemma}
\begin{Proof}
For any $f^s\in \ID$ with s an integer, there exists an integer k
such that $q+k(q-1) \geq s$. Then $f^sf^{q+k(q-1)-s}=f^{q+k(q-1)}
\in I$. By Lemma \ref{lm-2}, $f^{q+k(q-1)}
=f^qf^{k(q-1)}=f^{k(q-1)+1}=f^{q+(k-1)(q-1)}=\cdots =f^q=f.$ Thus,
we have $f\in I$, which implies that I is a radical ideal.
\end{Proof}

\begin{lemma} \label{lm-4}
Let $\ID$ be a polynomial  ideal in $\R_q$.
\begin{description}
\item[(1)] $\ID = (x_0-a_0,\ldots,x_{n}-a_{n})$ if and only if
$(a_0, \ldots,a_{n})$ is the only solution of $\ID$.
\item[(2)] $\ID = (1)$ if and only if $\ID$ has no solutions.
\end{description}
\end{lemma}
\begin{Proof}
If $\ID = (x_0-a_0,\ldots,x_{n}-a_{n})$, it is easy to see that
$(a_0, \ldots,a_{n})$ is the only solution of $\ID$.
Conversely, let $(a_0, \ldots,a_{n})$ be the only solution of $\ID$.
By Lemma \ref{lm-1}, we have $x_i-a_i=0$ on $\zero(\ID\cup\HS)$ in
$\F_q[\X]$, where $\HS$ is defined in \bref{H}. By Hilbert's
Nullstellensatz, there is an integer $s$ such that $(x_i-a_i)^s$ is
in the ideal generated by $\ID\cup\HS$ in $\F_q[\X]$. Considering
$\R_q$, it means that $(x_i-a_i)^s$ is in $\ID$. By Lemma
\ref{lm-3}, $\ID$ is a radical ideal in $\R_q$. Thus, $x_i-a_i$ is
in $\ID$. This prove (1).
For (2), if $\ID$ has no solution, we have
$\zero(\ID\cup\HS)=\emptyset$. By Hilbert's Nullstellensatz, $1\in
(\ID\cup \HS)$. That is, $1\in \ID$.
\end{Proof}

\begin{lemma} \label{lm-5}
Let $P \in \R_q$. $\hzero(P)=\F_q^n$ iff $P\equiv 0$.
$\hzero(P)=\emptyset$ iff
 $P^{q-1}-1\equiv 0.$
\end{lemma}
\begin{Proof}
If $P\equiv 0$, then $\hzero(P)=\F_q^n$. Conversely, we prove the
result by induction on $n$. If $n=1$, we consider the univariate
polynomial $P(x)\in \R_q$. Suppose that $P(x) \neq 0$. Since
deg$(P,x)\leq q-1$, $P$ has at most $q-1$ solutions in $\F_q$, a
contradiction. Now assume that the result has been proved for $n=k$.
For $n=k+1$, we have
$P(x_1,\ldots,x_n)=f_0x_n^{q-1}+f_1x_n^{q-2}+\cdots+f_{q-1}$, where
$f_i$ is a k-variable polynomial. By the induction hypothesis, if
some $f_i$ is not $0$, there exists an element
$(a_1,a_2,\ldots,a_{k})$ in $\F_q^k$ such that
$f_i(a_1,\ldots,a_k)\neq 0$. Then $P(a_1,\ldots,a_k)$ is a nonzero
polynomial whose degree in $x_{k+1}$ is less than $q$. Supposing
$a_{k+1}$ is not the solution of $P(a_1,\ldots,a_k)$,
$(a_1,\ldots,a_{k+1})$ is not
 the solution of $P$, a contradiction. Thus, we have $f_i=0$ for all $i$.
 It means that $P\equiv 0$, and the first result is proved.

If $\hzero(P)=\emptyset$, then $P\neq 0$ for any element in
$\F_q^n$, which implies that $P^{q-1}-1=0$ for any element in
$\F_q^n$. Then $P^{q-1}-1\equiv 0$. Conversely, suppose that there
is an element $\alpha\in \F_q^n$ such that $P(\alpha)=0$, which is
impossible since $P^{q-1}(\alpha)-1 \neq 0$. Thus,
$\hzero(P)=\emptyset$.
\end{Proof}

As a consequence of Lemma \ref{lm-5}, we have
\begin{cor}
Let $q=2$ and $P\in \R_2\setminus \F_2$. Then $\zero_2(P)\neq
\emptyset$.
\end{cor}

 But when $q>2$, the corollary is not correct. For example, considering
$\R_3$, it is easy to see that $\zero_3(x^2+1)=\emptyset$.

\begin{lemma}
\label{lm-41} Let $U,V$, and $D$ be polynomials  in $\R_q$. We have
\begin{eqnarray}
&&(U^{q-1}V^{q-1}-1) = (U^{q-1}-1,V^{q-1}-1).\label{eq-z1}\\
&&(U^{q-1}V^{q-1}-U^{q-1}-V^{q-1}) = (U,V).\label{eq-z2}\\
&&\hzero(UV) = \hzero(U)\cup \hzero(V).\label{eq-z3}\\
&&\hzero(\emptyset/D) = \hzero(D^{q-1}-1).\label{eq-z4}\\
&&\hzero(\PS)=\hzero(\PS\cup\{U\})\cup\hzero(\PS\cup\{U^{q-1}-
1\}).\label{eq-z5}
\end{eqnarray}
\end{lemma}
\begin{Proof} We have

$$\aligned
(U^{q-1}V^{q-1}-1)&=(U^{q-1}V^{q-1}-1,U^{q-1}(U^{q-1}V^{q-1}-1))\\
&=(U^{q-1}V^{q-1}-1,U^{q-1}V^{q-1}-U^{q-1})\\&=(U^{q-1}V^{q-1}-1,U^{q-1}-1)
=(U^{q-1}-1,V^{q-1}-1).\endaligned$$ This proves \bref{eq-z1}.
Equation \bref{eq-z2} can be proved similarly:
$$\aligned
(U^{q-1}V^{q-1}-U^{q-1}-V^{q-1})&=(U^{q-1}V^{q-1}-U^{q-1}-V^{q-1},U(U^{q-1}V^{q-1}-U^{q-1}-V^{q-1}))
\\&=(U^{q-1}V^{q-1}-U^{q-1}-V^{q-1},U)=(U,V).\endaligned$$
Since $\F_q$ is a field, \bref{eq-z3} is obvious. For any element
$\alpha \in \F_q^n, D(\alpha)\neq 0$ means that
$D^{q-1}(\alpha)-1=0$. Conversely, for any element $\alpha \in
\F_q^n,$ if $D(\alpha)=0$, we have $D^{q-1}(\alpha)-1\neq 0$. This
proves (6). Since $U(U^{q-1}-1)\equiv 0$, \bref{eq-z5} is a
consequence of \bref{eq-z3}.
\end{Proof}

 From (6) of Lemma~\ref{lm-41}, we can see that a quasi variety in
 $\F_q^n$ is also a variety.

\section{Proper Triangular Sets in $\R_q$}
\label{sec-ts}

In this section, we will introduce the concept of proper triangular
sets for which we can give an explicit formula for its number of
solutions.

\subsection{Triangular Sets}
\label{sec-asc}

Let  $P\in \R_q$. The {\bf  class} of $P$, denoted by $\class(P)$,
is the largest $c$ such that $x_c$ occurs in $P$. Then $x_c$ is
called the {\bf  leading variable} of $P$, denoted as $\lvar(P)$. If
$P\in \F_q$, we set $\class(P)=0$. If $\class(P)=c$, let us regard
$P$ as a univariate polynomial in $x_c$. We call ${\rm deg}(P,x_c)$
the {\bf  degree} of $P$, denoted as ${\rm deg}(P)$.
The coefficient of $P$ wrt $x_c^d$ is called the {\bf  initial} of
$P$, and is denoted by $\initial(P)$.
Then $P$ can be represented uniquely as the following form:
\begin{equation}\label{eq-p1} P = I x_c^d + U\end{equation}
where $I=\initial(P)$ and  $U$ is a polynomial with $\deg(U,x_c)<
d$. A polynomial $P_1$ has {\bf  higher ordering} than a polynomial
$P_2$, denoted as $P_2\prec P_1$, if $\class(P_1)>\class(P_2)$ or
$\class(P_1)=\class(P_2)$ and $\deg(P_1)>\deg(P_2)$.
If neither $P_1 \prec P_2$ nor $P_2 \prec P_1$, they are said to
have the same ordering, denoted as $P_1 \sim P_2 $.
It is easy to see that $\prec $ is a partial order on the
polynomials in $\R_q$.

A sequence of nonzero polynomials
\begin{equation}\label{eq-a}
\AS:~~ A_1, A_2,\ldots, A_r
\end{equation}
is a {\bf  triangular set} if either $r=1$ and $A_1\ne0$ or
$0<\cls(A_1) < \cdots < \cls(A_r)$.
A {\bf  trivial\/} triangular set is a polynomial set consisting
of a nonzero element in $\F_q$. For a triangular set $\A$, we denote
$\I_{\A}$ to be the product of the initials of the polynomials in
$\A$.
%
%

Let $\AS':~  A'_1, A'_2,\ldots, A'_{r'}$ and $\AS'':~ A_1'',
A_2'',\ldots, A_{r''}''$ be two triangular sets.  $\AS'$ is said to
be of {\bf  lower ordering\/} than $\AS''$, denoted as $\AS'\prec
\AS''$, if either there is some $k$ such that $A'_1\sim A_1'',
\ldots, A_{k-1}'\sim A_{k-1}''$,  while $A_k'\prec A_k''$; or
$r'>r''$ and $A'_1\sim A_1'',\ldots, A_{r''}'\sim A_{r''}''$.
We have the following basic property for triangular sets.

\begin{lemma}
\label{lm-a1}
A sequence of triangular sets steadily lower in ordering is
finite. More precisely, let $\A_1 \succ \A_2  \succ \cdots  \succ
\A_m$ be a strictly decreasing sequence of triangular sets in
$\R_q$. Then $m \le q^n$.
\end{lemma}
\begin{Proof}
Let $P$ be a polynomial in $\R_q$. If $\class(P)=c$ and $\deg(P)=d$,
$P$ and $x_c^d$ have the same ordering. Since we only consider the
ordering of the triangular sets, we may assume that the triangular
sets consist of powers of variables. In this case, two distinct
triangular sets can not have the same ordering. To form a triangular
set of this kind, we can choose one polynomial $M_i$ from
$\{0,x_i,x_i^2,\ldots,x_i^{q-1}\}$ for each $i$, and the triangular
set is $M_1,M_2,\ldots,M_n$. Note that when $M_i=0$, we will remove
it from the triangular set. Thus, there are $q^n-1$ nontrivial
triangular sets consist of powers of variables. Adding the trivial
triangular set consist of 1, we have a sequence of triangular sets
$\C_1 \succ \C_2 \succ \cdots \succ \C_{q^n}$. Let $\A_1 \succ \A_2
\succ \cdots \succ \A_m$ be a strictly decreasing sequence of
triangular sets. If $\A_i$ is nontrivial, for $P\in \A_i$, replace
it by $\lvar(P)^{\deg(P)}$. If $\A_i$ is trivial, replace it by $1$.
Then we get a strictly decreasing sequence of triangular sets $\B_1
\succ \B_2 \succ \cdots \succ \B_m$. This sequence must be a
sub-sequence of $\C_1 \succ \C_2 \succ \cdots \succ \C_{q^n}$.
Hence, $m\leq q^n$.
\end{Proof}

%

For two polynomials $P$ and $Q$, we use $\prem(Q,P)$ to denote the
pseudo-remainder of $Q$ with respect to $P$.
For a triangular set $\A$ defined in \bref{eq-a}, the  {\bf
pseudo-remainder} of $Q$ wrt $\A$ is defined recursively as
$$
\prem(Q,\A) = \prem(\prem(Q,A_r), A_1,\ldots,A_{r-1}) \hbox{ and }
\prem(Q,\emptyset)=Q.$$
Let $R = \prem(Q,\A)$. Then we have
\begin{equation}\label{eq-rf}
I_1^{s_1}I_2^{s_2}\cdots I_r^{s_r}Q = \sum_i Q_iA_i + R
\end{equation}
where $I_i=\initial(A_i)$ and $Q_i$ are some polynomials. The above
formula is called the {\bf  remainder formula\/}.
Let $\PS$ be a set of polynomials and $\A$ a triangular set. We
use $\prem(\PS,\A)$ to denote the set of nonzero $\prem(P,\A)$ for
$P \in \PS$.

A polynomial $Q$ is {\bf  reduced\/} wrt $P\ne0$ if $\cls(P)=c>0$
and $\deg(Q,x_c)<\deg(P)$. A polynomial $Q$ is {\bf  reduced\/} wrt
a triangular set $\A$ if $P$ is reduced wrt to all the polynomials
in $\A$. It is clear that the pseudo-remainder of any polynomial wrt
$\A$ is reduced wrt $\A$.

The {\bf  saturation ideal} of a triangular set $\A$ is defined as
follows
$$\sat(\A)=\{P\in \R_q|\,\, JP\in (\A)\}$$
where $J$ is a product of certain powers of the initials of the
polynomials in $\A$. We have

\begin{lemma} \label{lm-a2}
Let $\A=A_1,\ldots,A_r$ be a triangular set. Then
$\sat(\A)=(A_1,\ldots,A_r,\I_{\A}^{q-1}-1) $
\end{lemma}
\begin{Proof}
Denote $\ID=(A_1,\ldots,A_r,A_0)$ and $A_0=\I_{\A}^{q-1}-1$. If $P
\in \sat(\A)$, then $\I_{\A}^{q-1}P \in \A$. There exist polynomials
$B_i$ such that $\I_{\A}^{q-1}P=\sum_{i=1}^r B_iA_i.$ Hence,
$P=\sum_{i=1}^r B_iA_i-PA_0\in \ID$.
Conversely, let $P\in \ID$. Then there exist polynomials $C_i$ such
that $P=\sum_{i=1}^{r}C_iA_i+C_0A_0.$ Multiply $\I_{\A}$ to both
sides of the equation. Since $\I_{\A}(\I_{\A}^{q-1}-1)=0$, we have
$\I_{\A}P=\sum_{i=1}^r \I_{\A}C_iA_i.$ Thus, $P\in \sat(\A)$.
\end{Proof}

As shown by the following example, saturation ideals have
different properties comparing with that in the usual polynomial
ring.
\begin{example}\label{ex-21}
In $\R_3$, Let $\A = A_1,A_2$, $A_1=(x_1-1)x_2, A_2=(x_1+1)x_3$.
Then $\sat(\A)=(A_1,A_2,(x_1^2-1)^2-1)=(x_2,x_3,x_1)$.
\end{example}

\subsection{Proper Triangular Sets}

As we mentioned before, a triangular set could have no zero. For
example, $\zero_3(x^2+1)=\emptyset$.
To avoid this problem, we introduce the concept of proper triangular
sets.

A triangular set $\A=A_1,A_2,\ldots,A_r$ is called {\bf  proper}, if
the following condition holds: if $\class(A_i)=c_i$ and
$\deg(A_i)=d_i$, then $\prem(x_{c_i}^{q-d_i}A_i,\A)=0$.

The following lemmas show that proper triangular sets always have
solutions.

\begin{lemma} \label{lm-c1}
Let $P(x)$ be a univariate polynomial in $\R_q$, and suppose that
$\deg(P(x))=d$. If  prem$(x^{q-d}P(x),P(x))=0$, then $P(x)=0$ has
$d$ distinct solutions in $\F_q$.
\end{lemma}
\begin{Proof}
Since $P(x)$ is a univariate polynomial, $\initial(P) \in \F_q$.
If $\prem(x^{q-d}P(x),P(x))=0$ in $\R_q$, we have
$x^{q-d}P(x)=Q(x)P(x)$,
 where $Q(x)$ is a polynomial and $\deg(Q(x))<q-d$. Considering the above equation in $\F_q[x]$,
there is a polynomial $C$ such that
$x^{q-d}P(x)+C(x^q-x)=Q(x)P(x)$ in $\F_q[x]$, where  $x^{q-d}P(x)
+C(x^q-x)$ is equal to the canonical representation of
$\overline{x^{q-d}P(x)}$ in $\R_q$. Thus, we have
$(x^{q-d}-Q(x))P(x)=-C(x^q-x)$. Since all the elements of $\F_q$
are solutions of $x^q-x$, the $q$ distinct elements of $\F_q$ are
solutions of $(x^{q-d}-Q(x))P(x)$. Note that $\deg(Q(x))<q-d$.
Then $\deg(x^{q-d}-Q(x))=q-d$. Thus, $x^{q-d}-Q(x)$ has at most
$q-d$ solutions in $\F_q$, which means that $P(x)$ has at least
$d$ distinct solutions in $\F_q$. However, $\deg(P(x))=d$ implies
 $P(x)$ has at most $d$ solutions in $\F_q$. Hence, we can
conclude  $P(x)$ has $d$ distinct solutions in $\F_q$.
\end{Proof}

A  triangular set $\A$  is called {\bf  monic \/} if  the initial of
each polynomial in $\A$ is $1$. A monic  triangular set is of the
following form:
 $$A_1=x_{c_1}^{d_1}+U_1, A_2=x_{c_2}^{d_2}+U_2,\cdots, A_r=x_{c_r}^{d_r}+U_r$$
where $U_i$ is a polynomial in $x_1,\ldots,x_{c_i}$ such that
$\deg(U_i,x_{c_i}) < d_i$.

For a triangular set $\A:A_1,\ldots, A_r$, we call
$\deg(A_1)\deg(A_2)\cdots\deg(A_r)$ the {\bf  degree} of $\A$,
denoted as $\deg(\A)$. Let $\Y$ be the set $\{x_i\in \X|\,
\mbox{$x_i$ is the leading variable of some } A_j\in \A\}$. We use
$\U$ to denote $\X \setminus \Y$ and call the variables in $\U$ {\bf
parameters} of $\A$. Then we call $|\U|$ the {\bf  dimension} of
$\A$, denoted as $\dim(\A)$.

The following result shows that a monic  proper triangular set has
nice properties by giving an explicit formula for the number of
solutions. The result is useful because we will prove later that the
zero set for any polynomial system can be decomposed as the union of
the zero sets of monic  proper triangular sets.

\begin{theorem}\label{lm-monic}
Let $\A$ be a monic  triangular set. Then $\A$ is proper if and only
if $|\hzero(\A)|=\deg(\A)\cdot q^{\dim(\A)}$.
\end{theorem}
\begin{Proof}
Assume that $\A$ is proper. For the parameters in $\U$, we can
substitute them by any element of $\F_q$. Since $|\U|=\dim(\A)$,
there are $q^{\dim(\A)}$ parametric values for $\U$. For a
parametric value $U_0$ of $\U$ and a polynomial $P\in \R_q$, let
$P'$ denote $P(U_0)$. After the substitution, we obtain a new monic
triangular set $\A':A'_1,\ldots,A'_r$, where
$\class(A'_i)=\class(A_i)$ and $\deg(A'_i)=\deg(A_i)$. Let
$c_i=\class(A_i)$ and $d_i=\deg(A_i)$. Since $\A$ is a proper
triangular set, we have $x_{c_1}^{q-d_1}A_1=PA_1$. Then
$x_{c_1}^{q-d_1}A'_1=P'_1A'_1$. By Lemma \ref{lm-c1}, $A'_1$ has
$d_1$ distinct solutions. For a solution $\alpha$ of $A'_1$,
consider $A'_2(\alpha)$.
Since $\A$ is proper, we have $x_{c_2}^{q-d_2}A_2=Q_1A_1+Q_2A_2$
and hence
$x_{c_2}^{q-d_2}A'_2(\alpha)=Q'_1(\alpha)A'_1(\alpha)+Q'_2(\alpha)A'_2(\alpha)$.
Since $A'_1(\alpha)=0$, we have
$x_{c_2}^{q-d_2}A'_2(\alpha)=Q'_2(\alpha)A'_2(\alpha)$. By Lemma
\ref{lm-c1}, $A'_2(\alpha)$ has $d_2$ distinct solutions. By
repeating the process, we can prove that $\A'$ has $d_1d_2\cdots
d_r=\deg(\A)$ distinct solutions. Hence,
$|\hzero(\A)|=\deg(\A)\cdot q^{\dim(\A)}$.
%

Conversely, let us assume that $\A$ has $N=\deg(\A)\cdot
q^{\dim(\A)}$ solutions. Since $\A$ is monic, it means that for any
parametric value $U_0$ of $\U$ and any point $x$ in
$\hzero(A_1(U_0),\ldots,A_{i-1}(U_0))$, $A_i(U_0,x)$ has $\deg(A_i)$
distinct solutions.
Let $A_i=x_{c_i}^{d_i}+V_i$ for any $i$. For $A_1$, suppose
$\prem(x_{c_1}^{q-d_1}A_1,\A)=R_1\neq 0$. Then we have
$(x_{c_1}^{q-d_1}-P_1)A_1=R_1$, where $P_1$ is a polynomial. Choose
a parametric value $U_0$ of $\U$ such that $R_1(U_0)\neq 0$. Then
$A_1(U_0)$ has $d_1$ distinct solutions, this contradicts to
$0<\deg(R_1(U_0),x_{c_1})<d_1$. Thus, $R_1=0$. Now we consider
$A_2$. Suppose $\prem(x_{c_2}^{q-d_2}A_2,\A)=R_2\neq 0$. Then we
have two polynomials $Q_1$ and $Q_2$ such that
$x_{c_2}^{q-d_2}A_2=Q_1A_1+Q_2A_2+R_2$. Choose a parametric value
$U_1$ of $\U$ such that $R_2(U_1)\neq 0$. Since
$\deg(R_2,x_{c_1})<d_1$, there is a solution $x$ of $A_1(U_1)$ such
that $R_2(U_1,x)\neq 0$. Then we have
$(x_{c_2}^{q-d_2}-Q_1(U_1,x))A_2(U_1,x)=R_2(U_1,x)$. $A_2(U_1,x)$
has $d_2$ distinct solutions which contradicts to
$0<\deg(R_2(U_1,x_{c_2}))<d_2$. Thus, $R_2=0$. Similarly, we have
$\prem(x_{c_i}^{q-d_i}A_i,\A)=0$. Hence, $\A$ is proper.
\end{Proof}

As a consequence of Theorem \ref{lm-monic}, a monic proper
triangular set is square-free.

The concept of regular chains is important because of it has several
nice properties \cite{la1,rody1,yz1}.
The usual definition of regular chains need to be modified as shown
by the following example. This is due to the fact that $\R_q$ is a
ring with zero divisors.
\begin{example}
In $\R_3$, let $A_1=x_1x_2$, $A_1=(x_1^2-1)x_3$, and $\A =
A_1,A_2$. According to the usual definition, $\A$ is a regular
chain. $\A$ is also proper. But,
$\zero_3(\A/\I_\A)=\zero_3(\sat(\A))=\emptyset$ since $\I_\A =
x_1(x_1^2-1)=0$ in $\R_3$.
\end{example}

For two polynomials $P,Q\in \R_q$, let $\resl(P,Q,x_s)$ be the
resultant of $P$ and $Q$ wrt $x_s$
in $\R_q$. Let $\A$ be a triangular set of form \bref{eq-a} such
that $c_i = \cls(A_i)$. The resultant of $P$ wrt $\A$ is defined
recursively as: $\res(P,\A) = \res(\res(P,A_r,x_{c_r}),A_1,\ldots,$
$A_{r-1})$ and $\res(P,\{\})=P$.

A chain is called {\em regular} if
$$\prod_{i=1}^{n}\resl(I(A_i);A_1,\ldots,A_{i-1})\neq 0.$$
Regular chains have the following property.

\begin{theorem}
Let $\A$ be a regular and proper chain and $\U$ be the parameter set of
$\A$. Then, there exists a parametric value $U_0$ of $\U$ such that
$|\hzero(\A(U_0)/\I_{\A}(U_0))|=|\hzero(\A(U_0))|=\deg(\A)$.
%
\end{theorem}
\begin{Proof}
Let $R_i=\resl(I(A_i);A_1,\ldots,A_{i-1})$ and
$R=\prod_{i=1}^{n}R_i$. Since $R\neq 0$ and $R$ is a polynomial in
$\R_q[\U]$, by Lemma \ref{lm-5}, we can choose a parametric value
$U_0$ of $\U$ such that $R(U_0)\neq 0$. Then, we have $R_i(U_0)\neq
0$. $R_1(U_0)\neq 0$ means that $I_1(U_0)\neq 0$.
Similar to the proof of Theorem \ref{lm-monic}.
we can show that  $A_1(U_0)$ has $\deg(A_1)$ distinct solutions. $R_2(U_0)\neq 0$ implies that
$\hzero(I_2(U_0),A_1(U_0))=\emptyset.$ Thus, for a solution
$x_{1,1}$ of $A_1(U_0)=0$, $I_2(U_0,x_{1,1})\neq 0$ and
$A_2(U_0,x_{1,1})$ has $\deg(A_2)$ distinct solutions. Recursively,
we have $|\hzero(\A(U_0)/\I_{\A}(U_0))|=|\hzero(\A(U_0))|=\deg(\A)$.
\end{Proof}

\section{An Efficient Zero Decomposition Algorithm in $\R_q$}

In this section, we will give an improved algorithm which can be
used to decompose the zero set of a polynomial system into the union
of zero sets of monic triangular sets. Due to the special property
of $\R_q$, this algorithm has better properties and lower complexities than the general
zero decomposition algorithm and the output is stronger.

First, note that the following zero decomposition theorem
\cite{cade10,kalk1,la2,maza1,wang,wu-basic} is still valid and the
proof is also quite similar.
\begin{theorem}
\label{th-zdt} There is an algorithm which permits to determine for
a given polynomial  set $\PS$ in a finite number of steps regular and proper triangular
sets $\A_j,j=1,\ldots,s$ such that
 \[\hzero(\PS) = \cup_{j=1}^s \hzero(\A_j/\BI_{\A_j})= \cup_{j=1}^s \hzero(\sat(\A_j))\]
where $\sat(\A_j)$ is the saturation ideal of $\A_j$.
\end{theorem}

In $\R_q$, we can give the following improved zero decomposition
theorem which allows us to compute the number of solutions for a
finite set of polynomials.
\begin{theorem}\label{th-dzdt}  For a finite polynomial  set $\PS$, we can
compute monic proper triangular sets $\A_j,j=1,\ldots,s$ such that
\[\hzero(\PS) = \cup_{i=1}^s \hzero(\A_i)\]
such that $\hzero(\A_i)\cap \hzero(\A_j)=\emptyset$ for $i\ne j$. As
a consequence, we have
$$|\hzero(\PS)| = \sum_{i=1}^s \deg(\A_i)\cdot q^{\dim(\A_i)}.$$
\end{theorem}

\subsection{A Top-Down Characteristic Set Algorithm}
In this section, we will give a {\bf top-down characteristic set
algorithm {\bf TDCS}} that allows us to compute a decomposition
which has the properties mentioned in Theorem \ref{th-dzdt}.

Before giving the zero decomposition algorithm, we first give an
algorithm to compute a triangular set. The algorithm works from the
polynomials with the largest class and hence is a top-down zero
decomposition algorithm. The idea of top-down elimination is
explored in \cite{kapur1,wang}.
The key idea of the algorithm is as follows. Let $Q=Ix_c^d+U$ be a
polynomial with largest class and smallest degree in $x_c$ in a
polynomial set $\Q$.
If $I=1$, we can reduce the degrees of the polynomials in $\QS$ by
taking $\R=\prem(\Q,Q)$. Since $I=1$, we have
 $$\zero_q(\Q) = \zero_q(\R\cup\{Q\}).$$
If $I\ne1$, by \bref{eq-z5}, we split the zero set into two parts:
 \begin{equation}\label{eq-sp1} \zero_q(\Q) = \zero_q(\Q\cup\{I^{q-1}-1\})\cup
 \zero_q(\Q\setminus\{Q\}\cup\{I,U\}).\end{equation}
In the first part, since $I\ne0$ and $I^{q-1}-1=0$, $Q$ can be
replaced by $Q_1 =x_c^d +I^{q-2}U$ and we can treat this part as in
the first case.
The second part is simpler than $\Q$ and can be treated recursively.
The following {\bf well-ordering procedure} is based on the above
idea.

\begin{algorithm}{\bf---TDTriSet($\PS$)} \smallskip\\
   {\bf Input:} A finite set of polynomials $\PS$.\\
   {\bf Output:} A monic  triangular set $\A$ and a set of polynomial systems $\PS^*$
          such that $\hzero(\PS)=\hzero(\A)\cup_{\QS\in\PS^*}\hzero(\QS)$,
  $\hzero(\A)\cap \hzero(\QS_1)=\emptyset$, and $\hzero(\QS_1)\cap \hzero(\QS_2)=\emptyset$ for all $\QS_1,\QS_2\in\PS^*$.\medskip

 \noindent
1 Set $\A=\emptyset$ and $\PS^*=\emptyset$.\\
2 While $\PS\ne \emptyset$ do\\
  \SPC 2.1 If some nonzero element $\alpha$ of $\F_q$ is in  $\PS$, $\hzero(\PS)=\emptyset$. Return $\A=\emptyset$ and $\PS^*$.\\
  \SPC 2.2 Let $\PS_1\subset\PS$ be the polynomials with the highest class.\\
  \SPC 2.3 Let $Q\in \PS_1$ be a polynomial with lowest degree. \\
  \SPC 2.4 Let $Q=Ix_c^d+U$ such that $\class(Q)=c$, $\deg(Q)=d$ and $\initial(Q)=I$.\\
  \SPC 2.5 If $I=1$ do\\
  \SPC\SPC 2.5.1 Set $\R=\prem(\PS_1,Q)$.\\
  \SPC\SPC 2.5.2 If the classes of polynomials in $\R$ are lower
  than $c$\\
  \SPC\SPC\SPC\SPC(this situation will always happen when $q=2$), do\\
  \SPC\SPC\SPC\SPC\;  $\A = \A\cup\{Q\}$.\\
  \SPC\SPC\SPC\SPC\;  $\PS=\R\cup \{\PS\setminus
  \PS_1\}.$\\
  \SPC\SPC 2.5.3 Else, do\\
  \SPC\SPC\SPC\SPC\;
  $\PS=\R\cup\{Q\}\cup\{\PS\setminus \PS_1\} $ and goto 2.1.\\
   \SPC\ 2.6 Else do\\
  \SPC\SPC 2.6.1 Set $Q_1 = x_c^d +I^{q-2}U$ and $\PS_2 = \PS_1\setminus\{Q\}$.\\
  \SPC\SPC 2.6.2 $\PS=\prem(\PS_2,Q_1)\cup \{I^{q-1}-1\} \cup \{\PS\setminus \PS_1\}.$\\
  \SPC\SPC 2.6.3 $\PS_1=\{\PS\setminus \{Q\}\}\cup \A \cup
  \{I,U\}$.\\
  \SPC\SPC 2.6.4 $\PS^*=\PS^*\cup \{\PS_1\}$.\\
  \SPC\SPC 2.6.5 Set $\R=\prem(\PS_2,Q_1).$\\
  \SPC\SPC 2.6.6 If the classes of polynomials in $\R$ are lower
  than $c$, do\\
  \SPC\SPC\SPC\SPC\; $\A = \A\cup\{Q_1\}$.\\
  \SPC\SPC 2.6.7 Else, set $\PS=\PS\cup\{Q_1\}$ and goto 2.1.\\
3 Return $\A$ and $\PS^*$.\medskip
\end{algorithm}

The following theorem shows that to compute a monic  triangular set
in $\R_q$, we need only a polynomial number of polynomial arithmetic
operations. 

\begin{theorem}\label{th-tdzd}
Algorithm {\bf TDTriSet} is correct and in the whole algorithm we
need $O(n^2q^2+nlq)$ polynomial multiplications where $l = |\PS|$.
In particular, we need $O(nl)$ polynomial multiplications when
$q=2$.
\end{theorem}
\begin{Proof}
Let $\PS_1\subset\PS$ be the set of polynomials with the highest
class $c$ and $Q\in \PS_1$ a polynomial with lowest degree in $x_c$.
Let $c=\cls(Q)$, $d=\deg(Q)$ and $I=\initial(Q)$. If $I=1$, then
for $P\in \PS_1$, as a consequence of remainder formula
\bref{eq-rf}, $\hzero(\{Q,P\}) = \hzero(\{Q,\prem(P,Q)\})$.
Therefore, we have
$$\hzero(\PS) =\hzero((\PS\setminus \PS_1)\cup\{Q\}\cup\{\prem(P,Q)\ne0\,|\,
P\in\PS_1\}).$$
If $I\ne1$, by \bref{eq-z5}, we can split $\hzero(\PS)$ as the
following two parts:
\begin{eqnarray}
\hzero(\PS) &=&\hzero(\PS\cup\{I^{q-1}-1\})\cup\hzero(\PS\cup\{I\})\\
&=&\hzero((\PS\setminus \{Q\})\cup\{Q_1\}\cup\{I^{q-1}-1\})\cup
\hzero((\PS\setminus \{Q\})\cup\{I,U\})\label{split-2}
\end{eqnarray}
where $Q_1=x_c+I^{q-2}U$. The first part of \bref{split-2} can be
treated similarly to the case of $I=1$, and the second part of
\bref{split-2} will be a polynomial set in the output. This proves
that if we have the output it must be correct.

Now let us prove the termination of the algorithm. After each
iteration of the loop, the lowest degree of the polynomials with
highest class in $\PS$ will decrease. Then the highest class of
the polynomials in $\PS$ will be reduced and the polynomial $Q$
will be added to $\A$. Hence, the loop will end and give a
triangular set $\A$ and some polynomial sets $\PS^*$.

Finally, we will analyze the complexity of the algorithm.
Let $l = |\PS|$. After each iteration, the lowest degree of the
highest class of the polynomials in $\PS$ will be reduced at least
by one. Then, this loop will execute at most $n(q-1)$ times. After
each iteration, if $I=1$, then the new $\PS$ has at most $l$
polynomials. If $I\ne 1$, after this iteration there are two
cases:
\begin{itemize}
\item[(a)] Except $Q$ we still have some polynomials with this
class. Then, the new $\PS$ contains at most $l+1$ polynomials;

\item[(b)] The highest class is eliminated by $Q$. Then, the new
$\PS$ contains at most $l$ polynomials.
\end{itemize}
Therefore, in the whole algorithm there are at most $n(q-2)+l$
polynomials (The number is $l$ when $q=2$) .

In an iteration, suppose we use $Q=Ix_c^d+U$ to eliminate other
polynomials. First we should set $Q$ to be monic. It means that we
should compute $Q_1=x_c^d+I^{q-2}U$ and $I^{q-1}-1$, so we need
$2(q-2)$ polynomial multiplications. Thus, in the whole algorithm we
need at most $2n(q-1)(q-2)$ polynomial multiplications in order to
obtain the monic polynomials. Then we want to get $\prem(P,Q_1)$.
Since $Q_1$ is monic, it takes at most one polynomial multiplication
when we reduce the degree of $P$ by one. Let $D$ be the sum of the
degrees of polynomials with highest class. Then $D$ decreases by one
after one polynomial multiplication. Therefore, we need at most
$(n(q-2)+l)(q-1)-1$ multiplications to reduce $D$ from
$(n(q-2)+l)(q-1)$ to $1$. At the same time, we eliminate the highest
class. Thus, in the whole algorithm, we need at most
$n^2(q-2)(q-1)+nl(q-1)-n$ polynomial multiplications to get the
pseudo-remainders.
In all, the algorithm needs $O(n^2q^2+nlq)$ polynomial
multiplications, and when $q=2$ the number is $O(nl)$.
\end{Proof}

\begin{lemma}\label{lm-tdzd}
Let $\PS$ be an input of {\bf TDTriSet}. Assume that there is a
polynomial $P$ in $\PS$ such that $\class(P)=c$ and $\initial(P)=1$.
Let $\A$ be the monic triangular set in the output. Then, there is a
polynomial $P' \in \A$ such that $\class(P')=c$ and $\deg(P')\leq
\deg(P)$.
\end{lemma}

\begin{Proof}
Since there is a $P$ with class $c$, we need to deal with this
class. And we will eliminate this class by $P$ or by a $Q$ with
class $c$ and lower degree. This polynomial is the $P'$.
\end{Proof}

By using {\bf TDTriSet}, we have the following {\bf zero
decomposition algorithm}.
\begin{algorithm}\label{alg-tdzda}
   {\bf --- TDCS($\PS$)} \smallskip\\
  {\bf Input:} A finite set $\PS$ of polynomials.\\
  {\bf Output:} Monic  proper triangular sets satisfying the properties in Theorem \ref{th-dzdt}.\medskip

  \noindent
  1 Set $\PS^* = \{\PS\}$, $\A^* = \emptyset$ and $\C^*=\emptyset$.\\
  2 While $\PS^*\ne \emptyset$ do\\
  \SPC 2.1 Take a polynomial set $\QS$ from $\PS^*$ and set $\PS^*=\PS^*\setminus\{\QS\}$.\\
  \SPC 2.2  Let $\A$ and $\QS^*$ be the output  of {\bf TDTriSet} with input $\QS$. \\
  \SPC 2.3  if $\A \ne \emptyset$, set $\A^*=\A^* \cup \{\A\}$.\\
  \SPC 2.4 $\PS^*=\PS^* \cup \QS^*$\\
  3 Suppose $\A^*=\{\A_1,\ldots,\A_r\}$ and
  $\A_i=\{A_{i1},\ldots,A_{ip_i}\}.$\\
   4 Set $\PS^* = \{\}$ and for $i$ from 1 to $r$ do\\
  \SPC 4.1 Set $\B=\emptyset$.\\
  \SPC 4.2 For $j$ from 1 to $p_i$ do\\
  \SPC\SPC 4.2.1 Let $\class(A_{ij})=c_{ij}$ and $\deg(A_{ij})=d_{ij}$.\\
  \SPC\SPC 4.2.2
  If $R=\prem(x_{c_{ij}}^{q-d_{ij}}A_{ij},\A_i)\neq 0$, set $\B=\B\cup\{R\}$.\\
  \SPC 4.3 If $\B\neq \emptyset$, set $\PS^*=\PS^*\cup \{\A_i\cup \B\}.$\\
  \SPC 4.4 Else, set $\C^*=\C^*\cup\{\A_i\}$ \\
  5 If $\PS^*\neq \emptyset$,  set $\A^*=\emptyset$ and goto 2.\\
  6 Return $\C^*$\medskip
\end{algorithm}

\begin{theorem}
Algorithm {\bf TDCS} is correct.
\end{theorem}
\begin{Proof}
By Theorem \ref{th-tdzd}, if the loop in step 2 ends, we can obtain
$\A_1,\ldots,\A_q$ such that $\zero(\PS)=\cup_i\zero(\A_i)$.
In step 4, we check whether $\A_i$ is a proper triangular set. If it
is proper, we save it in the output list $\C^*$.
%
%
If $\A_i$ is not proper, suppose $\A_i=A_{i1},\ldots,A_{ip_i}$. we
add $\prem(x_{c_{ij}}^{q-d_{ij}}A_{ij},\A_i)\ne 0$ to $\A_i$, and
obtain a new polynomials set $\B_i$. We have
$\hzero(\A_i)=\hzero(\A_i,x_{c_{ij}}^{q-d_{ij}}A_{ij})=\hzero(\A_i,\prem(x_{c_{ij}}^{q-d_{ij}}A_{ij},\A_i))$.
Thus, $\hzero(\A_i)=\hzero(\B_i)$. Then we treated $\B_i$
recursively by step 2. Hence, if $\{\A'_1,\ldots,\A'_s\}$ is the output
of the algorithm, we have $\hzero(\PS)=\cup_i\hzero(A'_i)$.

Now we prove the termination of the algorithm. Firstly, we
prove the termination of step 2. For a polynomial set $\PS$, we
assign an index
$(c,c_{n,q-1},c_{n,q-2},\ldots,c_{n,1},\ldots,c_{1,q-1},\ldots,c_{1,1})$
where $c_{i,j}$ is the number of polynomials in $\PS$ and with class
$i$ and degree $j$ and for $i>c$, $\PS$ contains at most one polynomial with class
$i$ and this polynomial is monic. Note that, in the TDCS algorithm,
we need only to do eliminations on polynomials in $\PS$ with class smaller than
or equal to $c$.
To prove the termination of step 2, we will show that each polynomial
set in $\QS^*$ has a smaller index than that of $\QS$  in
the lexicographical ordering.
To prove this, we need only to show that in each step of Algorithm {\bf TDTriSet},
the updated polynomial set has a lower index than that of the original one.
In Algorithm {\bf TDTriSet}, the polynomial set $\PS$ is updated in three
ways.
Firstly, a polynomial $P$ is replaced by $\prem(P,Q)$ where $Q$ is a
monic polynomial. This will decrease of leading degree of $P$ and hence
decrease the index of the polynomial set.
Secondly, in step 2.6.2, the polynomial $Q$ is replaced by $Q_1$
and a new polynomial $I^{q-1}-1$ is added to the polynomial.
If $\prem(\PS_2,Q_1)\ne\emptyset$, the index of $\PS$ deceases since
the degrees of certain polynomials with class $c$ are decreased.
If $\prem(\PS_2,Q_1)=\emptyset$, the index of $\PS$ also deceases because
$Q_1$ is now the only polynomial with class $c$ in $\PS$ and the first
component in the index is deceased at least by one.
Thirdly,  in step 2.6.3, the polynomial $Q$ is replaced by $\{I,U\}$.
It is clear that the index of $\{I,U\}$ is less than the index of $\{Q\}$.
It is easy to show that a strictly decreasing sequence of indexes
must be finite. This proves the termination of the step 2.

Suppose we obtain $\A^*=\A_1,\ldots,\A_q$ after step 2. If all
$\A_i$ are proper, the algorithm will terminate. If
$\A_i=A_{i1},\ldots,A_{ip_i}$ is not proper, similar as above, we
obtain a polynomial set $\B_i$ such that there exist polynomials in
$\B_i$, which are reduced wrt $\A_i$.
To prove the termination of the whole algorithm, it is sufficient to
show that the new monic  triangular sets we obtain from $\B_i$ in
step 2 is of lower ordering than that of $\A_i$. Note that $\B_i
\setminus \A_i$ is the set of polynomials in $\B_i$ which are
reduced wrt $\A_i$.

Now let $\Q_1$ be the set of polynomials with highest class in $\B_i
\setminus \A_i$ and Q  be the one of lowest degree in $\Q_1$. Let
$Q=Ix_c^d+U$. Then in {\bf TDTriSet}, we splits $\hzero(\B_i)$ into
two parts:
$$
\hzero(\B_i)=\hzero(\{\B_i\setminus
\{Q\}\}\cup\{x_c^d+I^{q-2}U\}\cup\{I^{q-1}-1\})\cup
\hzero(\{\B_i\setminus\{Q\}\}\cup\{I,U\}).
$$
Note that $\A_i\subseteq \B_i$ and if there is a polynomial $A'$ in
$\A_i$ with class c then $deg(A')>deg(x_c^d+I^{q-2}U)$. Thus, by
Lemma \ref{lm-tdzd}, we can conclude that the monic triangular sets
we obtain from $\{\B_i\setminus
\{Q\}\}\cup\{x_c^d+I^{q-2}U\}\cup\{I^{q-1}-1\}$ is of lower ordering
than $\A_i$. For $\{\B_i\setminus\{Q\}\}\cup\{I,U\}$, it can be
recursively treated
 as $\B_i$. Hence, we prove the termination of the
algorithm.
\end{Proof}

We use the following simple example to illustrate how the algorithm
works.
\begin{example}
\label{ex4.1} In $\R_3$, let $\PS=\{x_1x_2x_3^2-1\}$.

In Algorithm {\bf TDTriSet},  we have
$\zero_3(\PS)=\zero_3(x_3^2-x_1x_2,x_1^2x_2^2-1)\cup\zero_3(x_1x_2,1)$.
Obviously, $\zero_3(x_1x_2,1)=\emptyset$. Then,
$\zero_3(\PS)=\zero_3(x_3^2-x_1x_2,x_1^2x_2^2-1)=\zero_3(x_3^2-x_1x_2,
x_2^2-1, x_1^2-1)\cup \zero_3(x_1^2,1)$. The algorithm returns
$\A=\{ x_1^2-1, x_2^2-1,x_3^2-x_1x_2\}$ and $\emptyset$.

In Algorithm {\bf TDCS},  we check whether $\A$ is proper:
$\prem(x_3(x_3^2-x_1x_2),\A)=(1-x_1x_2)x_3$,
$\prem(x_2(x_2^2-1),\A)=\prem(x_1(x_1^2-1),\A)=0$. We obtain a new
$\PS'=\{\A, (x_1x_2-1)x_3 \}$ such that
$\zero_3(\PS)=\zero_3(\PS')$.

Execute Algorithm {\bf TDTriSet} with input $\PS'$.
Choose $(x_1x_2-1)x_3$ to eliminate $x_3$. Then
$\zero_3(\PS')=\zero_3(x_3,x_3^2-x_1x_2,x_2^2-1,x_1x_2+1,x_1^2-1)
\cup \zero_3(x_3^2-x_1x_2,x_1x_2-1,x_2^2-1,x_1^2-1)$. For the first
part, we have
$\zero_3(x_3,x_3^2-x_1x_2,x_2^2-1,x_1x_2+1,x_1^2-1)=\zero_3(x_3,x_1x_2,x_2^2-1,x_1x_2+1,x_1^2-1)=\emptyset$.
 For the second part, we execute Algorithm {\bf TDTriSet} again and
 have
 $\zero_3(x_3^2-x_1x_2,x_1x_2-1,x_2^2-1,x_1^2-1)=\zero_3(x_3^2-x_1x_2,x_2-x_1,x_2^2-1,x_1^2-1)
    \cup
    \zero_3(x_3^2-x_1x_2,x_2^2-1,x_1^2-1,x_1,1)=\zero_3(x_3^2-x_1x_2,x_2-x_1,x_1^2-1)$.
   Let $\A'=\{x_3^2-x_1x_2,x_2-x_1,x_1^2-1\}$. Thus,
   $\zero_3(\PS)=\zero_3(\A')$.

 Returning to Algorithm {\bf TDCS},
   it is easy to check that $\A'$ is
   proper. Then we have
   $\zero_3(\PS)=\zero_3(x_3^2-1,x_2-x_1,x_1^2-1)$, and
   $|\zero_3(\PS)|=3^0(2\times1\times2)=4$.
\end{example}

%

\subsection{Complexity Analysis of {\bf TDCS} in $\R_2$}

As we mentioned in Section 1, a complexity analysis for the zero
decomposition algorithm is never given. Although, {\bf TDCS} is much
simpler than the zero decomposition algorithm over the field of
complex numbers, it is still too difficult to give a complexity
analysis. However, we are able to give a worst case complexity
analysis for algorithm {\bf TDCS} in the very important case of
$\R_2$.

In $\R_2$, it is easy to prove that a monic triangular set is always
proper. Therefore, we do not need to check whether a triangular set
is proper in Algorithm {\bf TDCS}.
Moreover, by \bref{eq-z2}, we can modify the Step 2.6.3 of {\bf
TDTriSet} as
 $$\PS_1=\{\PS\setminus \{Q\}\}\cup \A \cup
  \{U,I\}=\{\PS\setminus \{Q\}\}\cup \A \cup
  \{IU+I+U\},$$
  and call the new algorithm {\bf TDTriSet$_2$}.
After this modification, the number of polynomials in the new
component $\PS_1$ will not be bigger than $|\PS|$. From the proof of
Theorem \ref{th-tdzd}, we know that in the whole algorithm {\bf
TDTriSet$_2$} with input $\PS$ the number of polynomials is also at
most $|\PS|$. Then we obtain the following algorithm:
\begin{algorithm}\label{alg-tdzda2}
   {\bf --- TDCS$_2$($\PS$)} \smallskip\\
 {\bf Input:} A finite set of Boolean polynomials $\PS$.\\
  {\bf Output:} A sequence of monic triangular sets satisfying Theorem \ref{th-dzdt}.\medskip

  \noindent
  1 Set $\PS^* = \{\PS\}$, $\A^* = \emptyset$ and $\C^*=\emptyset$.\\
  2 While $\PS^*\ne \emptyset$ do\\
  \SPC 2.1 Choose a polynomial set $\QS$ from $\PS^*$.\\
  \SPC 2.2  Let $\QS$ be the input of {\bf TDTriSet$_2$}. Let $\A$ and $\QS^*$ be the output.\\
  \SPC 2.3  if $\A \ne \emptyset$, set $\A^*=\A^* \cup \{\A\}$.\\
  \SPC 2.4 $\PS^*=\PS^* \cup \QS^*$\\
  3 Return $\A^*$\medskip
\end{algorithm}

\begin{theorem}\label{th-bitsize}
The bitsize complexity of Algorithm {\bf TDCS$_2$} is $O(l^n)=O(2^{n
\log l})$, where $l$ is the number of polynomials in $\PS$.
\end{theorem}

\noindent{\bf Remark}. It is interesting to note that the complexity
for the exhaust search algorithm is $O(\|\PS\|\cdot2^n)$, where
$\|\PS\|$ is the bitsize of the polynomials in $\PS$ as defined in
Section 5.2. The complexity of the exhaust search is generally
better than our algorithm. But on the other hand, our algorithm can
solve nontrivial problems with $n\geq128$ as shown in Section 6.2
and Section 6.3, while it is clear that the exhaust search algorithm
cannot do that.
The complexity to compute a Gr\"obner basis of $\PS\cup\HS$ ($\HS$
is defined in \bref{H}) is known to be a polynomial in $d^n$ where
$d$ is the degree of the polynomials in $\PS$ \cite{la0}. Recently,
Bardet, Faugere, Salvy gave better complexity bounds under the
assumption of semi-regularity \cite{bardet1}.
It is an interesting problem that whether there exists a
deterministic algorithm to find all the solutions of a Boolean
polynomial system with complexity less than $O(2^n)$.

We will  prove Theorem \ref{th-bitsize} in the rest of this section.
In order to estimate the complexity of algorithm {\bf TDCS}$_2$, we
need to consider the worst case in the algorithm. We call the zero
decomposition process in the worst case {\bf W-Decomposition}.

In the worst case, we consider a set $\PS$ containing $l$ Boolean
polynomials which are with the highest class $n$ and the initials of
all these $l$ polynomials are not $1$. Then we need to choose one
polynomial $Q=Ix_n+U\in\PS$ and add $I+1$ to $\PS$. Let $Q_1 =
x_n+U$. Then we have:
\begin{equation}\label{eq-sp2}
\hzero(\PS)= \hzero(\prem(\PS\setminus \{Q\},Q_1),\cup \{Q_1,I+1\}))
\cup
 \hzero(\PS\setminus \{Q\}\cup\{IU+I+U\})
\end{equation}
In the worst case, we assume that the class of $I+1$ is $n-1$ and
$\prem(\PS\setminus \{Q\},Q_1)$ contains $l-1$ non-zero polynomials
with class $n-1$. Moreover, in the second  component in
\bref{eq-sp2}, we have a new polynomial $IU+I+U$ which is also of
class $n-1$. When we repeat the above procedure for the two
components in \bref{eq-sp2}, the above situations always happen. In
other words, in the worst case,when we eliminate a variable $x_c$,
the newly generated non-zero polynomials are always of class $c-1$.

We can illustrate the W-decomposition by the following figure:

\begin{tabular}{cccc}
 $(l,k,\ldots,\ldots)\Rightarrow$ &$(l-1,k+1,\ldots)\Rightarrow $&$(l-2,k+2,\ldots)\Rightarrow$ &$\cdots$  \\
 $\downarrow$ & $\quad\quad\downarrow$ &$\downarrow$ &  \\
              & $(0,l+k,\ldots)\Rightarrow \cdots$&$\vdots$ &   \\
 $(0,l+k,\ldots)\Rightarrow \cdots$ &$\downarrow$ & &  \\
 $\downarrow$ & $\vdots$ & & \\
 $\vdots$&  &  & \\
\end{tabular}

In this figure and the rest of this section,
$(l_n,l_{n-1},\cdots,l_1)$ represents a polynomial set which
contains $l_i$ polynomials with class $i$. The right arrows point to
the second component in \bref{eq-sp2}, while the down arrows point
to the first component in \bref{eq-sp2} or more precisely, to
$\prem(\PS\setminus \{Q\},Q_1)\cup \{I+1\}$.

To solve a polynomial set $\PS$ with $l$ elements,  we will obtain a
lot of components.
We can sort these components into $n$ groups by the variables
involved in them. For any $i=1,2,\ldots,n$, the i-th group consists
of the components where the variables to be eliminated are
$\{x_1,x_2,\ldots,x_i\}$.
Suppose there are $k_i$ elements in the i-th group. We define the
{\bf time-polynomial } of $\PS$ to be
 \begin{equation}\label{eq-sp3}
 B(\PS)=k_nT_n+k_{n-1}T_{n-1}+\cdots+k_1T_1\end{equation}
where $T_i$ is a quantity to measure the complexity for executing
{\bf TDTriSet$_2$} whose input is a polynomial set consisting of $l$
polynomials in $i$ variables $\{x_1,x_2,\ldots,x_i\}$. $T_i$ could
be the bitsize of the involving polynomials or the number of
arithmetic operations needed in the algorithm.
Obviously, $B(\PS)$ gives the corresponding worst case complexity
when the meaning of $T_i$ is fixed.

For two polynomial sets $\PS_1$ and $\PS_2$, let
$B(\PS_1)=k_nT_n+\cdots+k_1T_1$ and $B(\PS_2) =
k'_nT_n+\cdots+k'_1T_1$. If $k_i>k'_i$ for all $i$, we say that
$B(\PS_1)$ is of {\bf higher ordering} than $B(\PS_2)$, denoted by
$B(\PS_1)>B(\PS_2)$. We define
 $$S(\PS)=B(\PS)-T_c$$
  where $c$ is the highest class of the polynomials in $\PS$. Thus, $S(\PS)$
is the complexity for solving all the components which are
originated from the second component in \bref{eq-sp2}. The order of
$S(\PS)$ can also be defined as $B(\PS)$. Therefore, we can use
equation \bref{eq-sp3} as the recursive formula to compute the worst
case complexity of the algorithm.

The following result shows that the problems solved with
w-decomposition is indeed the worst case in terms of complexity.

\begin{lemma}\label{lm-tpoly}
Let $\QS$ be a polynomial set of the form $(l,0,\ldots,0)$, which
need to be solved with w-decomposition.
Let $B(\PS)$ be the time-polynomial of any other problem with
$|\PS|\leq l$. We have $B(\QS)\geq B(\PS)$ and $S(\QS)\geq S(\PS)$.
\end{lemma}
\begin{Proof}
We prove the lemma by induction. If $n=1$, no components are
generated, so we have $B(\PS)=T_1$ and $S(\PS)=0$ for any problem,
and the lemma holds for $n=1$. Now suppose we have proved the lemma
for $n=k$. If $n=k+1$, we have the following figure for the
w-decomposition of problem $(l,0,\ldots,0)$:

\begin{tabular}{cclcccc}
$(l,0,\ldots,0)\Rightarrow$ &$(l-1,1,\ldots,0)\Rightarrow$ &$\cdots$&$\Rightarrow$& $(1,l-1,0,\ldots,0)\Rightarrow$&$(0,l,0,\ldots,0) $\\
$\downarrow$ & $\downarrow$ &  & &$\downarrow$&  \\
$(0,l,0,\ldots,0)$  &$(0,l,0,\ldots,0)$ &$\cdots$&&$(0,l,0,\ldots,0)$&  \\
\end{tabular}

\noindent We can get the following recursive formula for the
time-polynomial of $(l,0,\ldots,0)$:

\begin{equation}\label{eq-time}
B(l,0,\ldots)=lT_n+B(0,l,0,\ldots)+lS(0,l,0,\ldots,0)
\end{equation}
where $(0,l,0,\ldots)$ represents a w-decomposition problem with $l$
input polynomials in variable $\{x_1,\ldots,x_{n-1}\}$

For any other polynomial set $\PS$ with no more than $l$ input
polynomials, we can write it as $(l_n,l_{n-1},\ldots,l_1)$. If
$l_n=0$ the lemma can be proved easily from equation
(\ref{eq-time}). Now we assume $l_n>0$. For the $l_n$ polynomials
with class $n$, if there is a polynomial with initial $1$, we will
not generate any component when we eliminate class $n$, then
$B(\PS)=T_n+S(\PS')$. Note that $|\PS'|\leq l$ and the elements of
$\PS'$ are all have $n-1$ variables $\{x_1,\ldots,x_{n-1}\}$. Thus
$B(l,0,\ldots)\geq B(\PS)$ and $S(l,0,\ldots)\geq S(\PS)$ by the
hypothesis.

If there exist no polynomials with initial $1$ in these $l_n$
polynomials. we have the the following decomposition figure:
\begin{center}
\begin{tabular}{cclcccc}
$(l_n,\ldots)\Rightarrow$ &$(l_n-1,\ldots)\Rightarrow$ &$\cdots$&$\Rightarrow$& $(1,\ldots)\Rightarrow$&$\PS_0 $\\
$\downarrow$ & $\downarrow$ &  & &$\downarrow$&  \\
$\PS_1$  &$\PS_2$ &$\cdots$&&$\PS_{l_n}$&  \\
\end{tabular}
\end{center}
Thus, we have $$B(\PS)=l_nT_n+B(\PS_0)+\sum_{i=1}^{l_n}S(\PS_i).$$
Note that $\PS_i$ has at most $n-1$ variables $\{x_1,\ldots,x_n\}$
and $|\PS_i|\leq l$, for any $i=0,1,\ldots,l_n$. By the hypothesis
we have $S(\PS_i)\leq S(0,l,0,\ldots,0)$ and $B(\PS_0)\leq
B(0,l,0,\ldots,0)$. Since $l\geq l_n$ we can conclude that
$B(l,0,\ldots)\geq B(\PS)$ and $S(l,0,\ldots)\geq S(\PS)$.
Consequently, the lemma holds in any case for $n=k+1$.
\end{Proof}

{\em Proof of Theorem \ref{th-bitsize}}.
From equation (\ref{eq-time}), we can obtain the value of
$B(l,0,\ldots,0)$. Write $B(0,\ldots,0,l,0,\ldots,0)$ as $B_i$ and
$S(0,\ldots,0,l,0,\ldots,0)$ as $S_i$, where $l$ is in the i-th
coordinate. Then we have $B_n=l(T_n-T_{n-1})+(l+1)B_{n-1}$. It is
easy to check that for $n\geq 3$ we have
$$B_n=lT_n+l^2T_{n-1}+l^2(l+1)T_{n-2}+\cdots+l^2(l+1)^{n-3}T_2+(l+1)^{n-2}T_1.$$
If the variables of input polynomials are $\{x_1,\ldots,x_k\}$, the
number of monomials occuring in {\bf TDTriSet$_2$} are at most
$2^k$, and therefore the bitsize complexity of multiplication is
$2\cdot4^k$. By Theorem \ref{th-tdzd}, we can substitute $T_k$ with
$(2\cdot4^k)k(l-1)$ for any $k\geq 2$ and $T_1$ can be set to $0$.
We have $B_n\thickapprox
2(4^3l^{n+1}-4^{n+1}l^3)/(l-4)^2+4^3l(l^n-2nl4^{n-2})/(l-4)$. Since
$l>>4$, we have proved Theorem \ref{th-bitsize}.

\section{A Multiplication Free Zero Decomposition Algorithm in $R_2$}
\label{sec-mf}

It is known that a major difficulty in computing a zero
decomposition is the occurrence of large polynomials
which are caused mainly by multiplication of polynomials.
Due to this reason, even the procedure to compute one triangular set,
called well-ordering procedure in \cite{wu-basic}, has exponential complexity
for all known CS methods.
In order to overcome this difficulty, we introduce a zero decomposition
algorithm in $\R_2$, where only additions of polynomials are used.
We show that the well-ordering procedure in our multiplication free
algorithm has polynomial time complexity for input polynomials with fixed degree.

\subsection{The Algorithm}
The key idea of the algorithm is to avoid polynomial multiplications.
Before doing the pseudo remainders, we reduce the initials of the
polynomials in $\PS_1$ in step 2.2 of the Algorithm {\bf TDTriSet}
to 1 by repeatedly using \bref{eq-sp1}. For such polynomials, we
have the following result.
\begin{lemma}\label{lm-prem1}
Let $P=x_c+U_1$ and $Q=x_c+U_2$ be polynomials with class $c$ and
initial 1. Then, we have $\deg(\prem(Q,P)) \le
\max\{\deg(U_1),\deg(U_2)\}$.
\end{lemma}
\begin{Proof}
 In that case, the pseudo-remainder needs additions
only:  $\prem(Q,P) = U_1+U_2$. The lemma follows from this formula
directly.
\end{Proof}

Based on the above idea, Algorithm {\bf TDTriSet} can be modified to
the following multiplication free (MF) {well-ordering procedure}
to compute a triangular set.

\begin{algorithm}\label{alg-wop3} {\bf --- MFTriSet($\PS$)}

\noindent {\bf Input:}A finite set of polynomials $\PS$.\\
  {\bf Output:} A monic  triangular set $\A$ and a set of polynomial systems $\PS^*$
          such that $\tzero(\PS)=\tzero(\A)\cup_{\QS\in\PS^*}\tzero(\QS)$,
  $\tzero(\A)\cap \tzero(\QS_1)=\emptyset$, and $\tzero(\QS_1)\cap \tzero(\QS_2)=\emptyset$ for all $\QS_1,\QS_2\in\PS^*$.
  \medskip

  \noindent
  1 Set $\PS^* = \{\}$, $\A = \emptyset$.\\
  2 While $\PS\ne \emptyset$ do\\
  \SPC 2.1 If $1\in\PS$, $\tzero(\PS)=\emptyset$. Set $\A=\emptyset$ and return $\A$ and $\PS^*$.\\
  \SPC 2.2 Let $\PS_1\subset\PS$ be the polynomials with the highest class.\\
  \SPC 2.3 Let $\PS_2=\emptyset$, $\QS_1 = \PS\setminus\PS_1$.\\
  \SPC 2.4 While $\PS_1\ne \emptyset$ do\\
  \SPC\SPC\SPC  Let $P=Ix_c+U\in \PS_1$, $\PS_1=\PS_1\setminus\{P\}$.\\
  \SPC\SPC\SPC  $\QS_2 = \PS_1\cup\QS_1\cup\PS_2\cup\{I,U\}$.\\
  \SPC\SPC\SPC  $\PS^*= \PS^*\cup \{\QS_2\}$.\\
  \SPC\SPC\SPC  $\PS_2=\PS_2\cup\{x_c+U\}$, $\QS_1=\QS_1\cup\{I+1\}$.\\
  \SPC 2.5 Let $Q=x_c+U$ be a polynomial with lowest degree in $\PS_2$.\\
  \SPC 2.6 $\A = \A\cup\{Q\}$.\\
  \SPC 2.7 $\PS = \QS_1\cup\prem(\PS_2,Q)$.\\
  3 Return $\A$ and $\PS^*$.
  \medskip
\end{algorithm}

In Step 2.4, we use formula \bref{eq-sp1} in $\R_2$, that is, for $P=Ix_c+U$,
 $$\tzero(P) =  \tzero(\{x_c+U, I+1\}) \cup \tzero(\{I,U\})$$
to split the polynomial set.

With Algorithm {\bf MFTriSet}, we can easily give a
multiplication-free zero decomposition algorithm: we just need to
replace Algorithm {\bf TDTriSet$_2$} by Algorithm {\bf MFTriSet} in
Algorithm {\bf TDCS$_2$}. We call this algorithm {\bf MFCS}.

\begin{algorithm}\label{alg-mfzda}
   {\bf --- MFCS($\PS$)} \smallskip\\
   {\bf Input:} A finite set of polynomials $\PS$.\\
   {\bf Output:} Monic  proper triangular sets satisfying the properties in Theorem \ref{th-dzdt}.\medskip

  \noindent
  1 Set $\PS^* = \{\PS\}$, $\A^* = \emptyset$ and $\C^*=\emptyset$.\\
  2 While $\PS^*\ne \emptyset$ do\\
  \SPC 2.1 Choose a polynomial set $\QS$ from $\PS^*$.\\
  \SPC 2.2  Let $\QS$ be the input of {\bf MFTriSet}. Let $\A$ and $\QS^*$ be the output.\\
  \SPC 2.3  if $\A \ne \emptyset$, set $\A^*=\A^* \cup \{\A\}$.\\
  \SPC 2.4 $\PS^*=\PS^* \cup \QS^*$\\
  3 Return $\A^*$\medskip
\end{algorithm}


{\bf Remark}. In the following, we will analyze the complexity of
Algorithm {\bf MFTriSet}. Basically, we will show that the size of
the polynomials in bounded by the size of the input polynomials and
the worst case complexity of this algorithm is roughly $O(n^d)$.
The second result implies that for a fixed $d$, say $d=2$, Algorithm
{\bf MFTriSet} is a polynomial time algorithm. Note that solving
quadratic Boolean equations is NP complete. In Algorithm {\bf MFCS},
the number branches could be exponential. We will discuss how to control the
number of branches in Section 6.

\subsection{Bitsize Bounds of the Polynomials in {\bf MFTriSet}}
\label{sec-mf2}

 In order to estimate the size of the polynomials, we
introduce a {\bf bitsize measure} for a polynomial in $\R_2$.
Let $M=x_{i_1}x_{i_2}\cdots x_{i_k}$ be a monomial. The length of
$M$, denoted by $\|M\|$, is defined to be k. Specially, the length
of $1$ is defined as $1$. For a polynomial $P=M_1+\cdots+M_t$ where
$M_i$ are monomials, $\|P\|=\sum_{i=1}^t\|M_i\|$ is called the {\bf
length} of $P$.

We first note that since Algorithm {\bf MFCS} is multiplication
free, the degrees of the polynomials occurring in the algorithm will
be bounded by $d=\max_{P\in\PS}\{\deg(P)\}$. As a consequence, the
size of the polynomials occurring in the algorithm will be bounded
by  $O(n^d)$. Then, the size of the polynomials is effectively
controlled if $d$ is small. For all the examples in Section 6, we
have $d\le 4$ and $n$ ranges from $40$ to $128$.
For such examples, the polynomials have size $O(n^4)$, while the
largest possible polynomial in $n$ variables has size $O(2^n)$.

In the following theorem, we will further show that the size of the
polynomials in Algorithm {\bf MFTriSet} are effectively controlled
in all cases.

\begin{theorem}\label{th-len1}
Let $n$ be the number of variables and $\PS$ the input of Algorithm
{\bf MFTriSet}.
Then, for any polynomial $T$ occurring in Algorithm {\bf MFTriSet},
we have $\|T\|\leq \sum_{P\in\PS}\|P\|$.
If $|\PS|>n$, then there exist $n$ polynomials $P_1,\ldots, P_n$ in
$\PS$ such that $\|T\|\leq \|P_1\|+\|P_2\|+\cdots+\|P_n\|$.
\end{theorem}

This result is nontrivial, because repeated additions of polynomials
can increase the size of the polynomials by an exponential factor.
The proof of this result is quite complicated. Intuitively, we want
to show that a polynomial $P$ used in early steps of the algorithm
will be ``canceled" in later steps by addition of two polynomials
both containing $P$, that is, $(P_1+P) + (P_2+P) = P_1+P_2$.


In order to prove Theorem \ref{th-len1}, we need to prove several
lemmas first.
Let $k$ be an integer and $P$ be a polynomial. Write $P=Ix_k+U$ as a
univariate polynomial in $x_k$. We define two operators $\OR_k$ and
$\OJ_k$ as follows:
 \begin{equation}\label{eq-rj}
  \OR_k(P)=U, \OJ_k(P)=I+1 \hbox{ if }\cls(P)=k.\quad
  \OR_k(P)=P, \OJ_k(P)=0 \hbox{ if } \cls(P)<k.
 \end{equation}
Then, we have the following lemma
\begin{lemma}\label{lm-o1}
Let $P$ and $Q$ be polynomials with $\cls(P)\leq k$ and
$\cls(Q)\leq k$. Then
\begin{itemize}
\item[(1)] $\OR_k(P+Q)=\OR_k(P)+\OR_k(Q)$;

\item[(2)] $\OR_k(P+1)=\OR_k(P)+1$;

\item[(3)]
If $\cls(P)=\cls(Q)=k$ then $\OJ_k(P+Q)=\OJ_k(P)+\OJ_k(Q)+1$;
otherwise $\OJ_k(P+Q)=\OJ_k(P)+\OJ_k(Q)$.
\end{itemize}
\end{lemma}

\begin{Proof}
It is easy to check.
\end{Proof}

Note that we can define the composition  of $\OR$ and $\OJ$
naturally. Let $\OS_{j,k}=\{\OO_j\OO_{j+1}\ldots \OO_k|\;$
$\OO_i=\OR_i \mbox{ or } \OJ_i,\; i=j,\ldots,k \}$, where $ 1\leq
j\leq k\leq n.$
\begin{lemma}\label{lm-o2}
Let $P$ be a polynomial with $\cls(P)=k$. Then $\sum_{L_{j,i}\in
\OS_{j,k}}\|L_{j,i}P\|\leq \|P\|$ for any fixed $j=1,2,\ldots,k$.
\end{lemma}

\begin{Proof}
For a polynomial $Q=Ix_c+U$ with $I \ne 1$,  we have
$\|Q\|\geq\|I\|+|U\|+1$. $\OJ_cQ=I+1$ and $\OR_cQ=U$. Therefore,
$\|\OJ_cQ\|+\|\OR_cQ\|=\|I+1\|+\|U\|\leq \|I\|+\|U\|+1\leq\|Q\|$. If
$I=1$, we have  $\|\OJ_cQ\|+\|\OR_cQ\|=0+\|U\|<\|Q\|$. For $i>c$, we
have $\OJ_iQ=0$ and $\OR_iQ=Q$. Then $\|\OJ_iQ\|+\|\OR_iQ\|=\|Q\|$.
Hence, in any case, we have $|\OJ_iQ\|+\|\OR_iQ\|\leq \|Q\|$.

For any $j$, we have
$\sum_{L_{j,i}\in\OS_{j,k}}\|L_{j,i}P\|=\sum_{L_{j+1,i}\in
\OS_{j+1,k}} (\|\OJ_jL_{j+1,i}P\|+\|\OR_jL_{j+1,i}P\|) \leq
\sum_{L_{j+1,i}\in \OS_{j+1,k}}\|L_{j+1,i}P\|\leq \cdots \leq
\|\OJ_kP\|+\|\OR_kP\|\leq \|P\|.$
\end{Proof}

%


\noindent{\bf  Proof of Theorem \ref{th-len1}}:
 For any $k=1,\ldots, n$, we assume that in the $k$-th
round  of {\bf MFTriSet} we deal with the polynomials of class $k$.
In algorithm {\bf MFTriSet}, when we compute the pseudo-remainder of
two polynomials $P$ and $Q$ in the $k$-th round, we set their
initials to 1 at first, and then compute a new polynomial
$\OR_kP+\OR_kQ$. Thus,  a polynomial $P^{(k)}$ in $k$-th round can
be obtained in three ways:
\begin{itemize}

\item[(1)] $P^{(k)}$ is an input polynomial;

\item[(2)] $P^{(k)}=\initial(Q^{(k+i)})+1$ for some $Q^{(k+i)}$ of
round $k+i$.
$P^{(k)}=\OR_{k+1}\cdots\OR_{k+i-1}\OJ_{k+i}Q^{(k+i)}$.

\item[(3)] $P^{(k)}=\OR_{k+j}(Q_1^{(k+j)}+Q_2^{(k+j)})=
\OR_{k+1}\cdots\OR_{k+j}(Q_1^{(k+j)}+Q_2^{(k+j)})=\OR_{k+1}\cdots\OR_{k+j}Q_1^{(k+j)}+\OR_{k+1}\cdots\OR_{k+j}Q_2^{(k+j)}$,
where $Q_1^{(k+j)}$ and $Q_2^{(k+j)}$ are polynomials of round
$k+j$.
\end{itemize}
In the cases 2 and 3, if $i$ and $j$ are bigger than $1$, we still
regard $\OR_{k+2}\cdots\OR_{k+i-1}\OJ_{k+i}Q^{(k+i)}$,
$\OR_{k+2}\cdots\OR_{k+j}Q_1^{(k+j)}$ and
$\OR_{k+2}\cdots\OR_{k+j}Q_2^{(k+j)}$ as polynomials of round $k+1$.
In this way, we can represent $P^{(k)}$ by operators and polynomials
of round $k+1$. We call it the {\bf  backtracking} representation of
$P^{(k)}$.  Now we can consider these polynomials of round $k+1$ and
get the backtracking
 representation of them. By Lemma \ref{lm-o1}, we can get a
representation of $P^{(k)}$ by composite operators and polynomials
in round $k+2$. Then, we can do the process recursively. In the
process of computing the backtracking representation, when meet an
input polynomial, we stop representing this polynomial by the ones
of higher round. At last, we backtrack to the round $n$, and
eliminate the terms composed of the same operators and polynomials.
Note that the polynomials of round $n$ are all from the input. Then
 we have
\begin{equation}\label{eq-o1}
 P^{(k)}=\sum_{i=1}^{r_n}\sum_{L_j\in T_{n,i}}L_jQ_i^{(n)}+\sum_{i=1}^{r_{n-1}}\sum_{L_j\in
 T_{n-1,i}}L_jQ_i^{(n-1)}+\cdots+\sum_{i=1}^{r_{k+1}}\sum_{L_j\in
 T_{k+1,i}}L_jQ_i^{(k+1)}
\end{equation}
or
\begin{equation}\label{eq-o2}
P^{(k)}=\sum_{i=1}^{r_n}\sum_{L_j\in
T_{n,i}}L_jQ_i^{(n)}+\sum_{i=1}^{r_{n-1}}\sum_{L_j\in
 T_{n-1,i}}L_jQ_i^{(n-1)}+\cdots+\sum_{i=1}^{r_{k+1}}\sum_{L_j\in
 T_{k+1,i}}L_jQ_i^{(k+1)}+1
\end{equation}
where $T_{m,i}\subseteq \OS_{k+1,m}$ is a set of composite operators
and $Q_i^{(m)}$ is an input polynomial with class $m$
($m=k+1,\ldots,n, i=1,\ldots,r_m$). The appearance of $1$ is due to
the equation (3) of Lemma \ref{lm-o1}.  The number of different
polynomials in the above equation, denoted by $N$, is
$r_{k+1}+r_{k+2}+\cdots+r_n$.

Now we will give an upper bound for $N$. It is easy to see that,
when we backtrack to the round $k+1$, there exist at most two
different polynomials. Suppose that now we backtrack to the round
$k+i$, and there are $t$ different polynomials in the
representation. Then, $t_1$ of them are the form of $\OR_{k+i+1}f$,
where $f$ is a polynomial with $\cls(f)<k+i+1$; $t_2$ of them are
the form of $\OJ_{k+i+1}g$, where $\cls(g)=k+i+1$; $t_3$ of them are
input polynomials. Thus, the others can be represented as
$\OR_{k+i+1}h+\OR_{k+i+1}h_i$, where $h$ is a fixed polynomial with
$\cls(h)= k+i+1$ and $h_i$ is some polynomial with
$\cls(h_i)=k+i+1$. Therefore, the number of different polynomials in
the representation of round $k+i+1$ is at most
$2(t-t_1-t_2-t_3)-(t-t_1-t_2-t_3-1)+t_1+t_2+t_3=t+1$. Hence, when we
backtrack to the round $n$, we have $N \leq  n-k+1 $.

 For any $m=k+1,\ldots,n,\;
i=1,\ldots,r_m$, since $T_{m,i}\subseteq \OS_{k+1,m}$, by Lemma
\ref{lm-o2}, we have $\sum_{L_j\in T_{m,i}}\|L_jQ_i^{(m)}\|\leq
\sum_{L_j \in \OS_{k+1,m}}\|L_jQ_i^{(m)}\| \leq \|Q_i^{(m)}\|$.
\begin{itemize}

\item[(a)] Suppose that  $P^{(k)}$ is of form
\bref{eq-o1}. We have $\|P^{(k)}\|\leq
\sum_{m=k+1}^n\sum_{i=1}^{r_m} \|Q_i^{(m)}\|$ where
$r_{k+1}+\cdots+r_{n}\leq n-k+1 \leq n$.

\item[(b)] Suppose the representation of $P^{(k)}$ is equation
\bref{eq-o2}. It is easy to see that there exists a term of the form
$\OR_{k+1}\cdots\OR_{k+i-1}\OJ_{k+i}LQ^{(k+j)}$, where $Q^{(k+j)}$
is an input polynomial with class $k+j$, $L\in \OS_{k+i+1,k+j}$ and
$\cls(LQ^{(k+j)})=k+i$. If $\initial(LQ^{(k+j)})= W+1$ where $W$ is
a polynomial without a constant term, we have
$\OJ_{k+i}LQ^{(k+j)}=W$. Therefore
$\|\OJ_{k+i}LQ^{(k+j)}\|+\|\OR_{k+i}LQ^{(k+j)}\|<\|LQ^{(k+j)}\|$.
Hence, $\|P^{(k)}\|<\sum_{m=k+1}^n\sum_{i=1}^{r_m}\|Q_i^{(m)}\|+1$
which means $\|P^{(k)}\|\leq
\sum_{m=k+1}^n\sum_{i=1}^{r_m}\|Q_i^{(m)}\|$. If
$\initial(LQ^{(k+j)})\\=W$ where $W$ is a polynomial without a
constant term, we have $\OJ_{k+i}LQ^{(k+j)}=W+1$. Thus,
$P^{(k)}=\OR_{k+1}\cdots\OR_{k+i-1}\OJ_{k+i}LQ^{(k+j)}+1+E=\OR_{k+1}\cdots\OR_{k+i-1}W+E$
where $E$ is the sum of other terms in equation \bref{eq-o2}.
Obviously,
$\|\OR_{k+1}\cdots\OR_{k+i-1}W\|<\|\OR_{k+1}\cdots\OR_{k+i-1}(W+1)\|
=\|\OR_{k+1}\cdots\OR_{k+i-1}\OJ_{k+i}LQ^{(k+j)}\|$. Then $\|P\|<
\|\OR_{k+1}\cdots$ $\OR_{k+i-1}\OJ_{k+i}LQ^{(k+j)}\|+\|E\|\leq
\sum_{m=k+1}^n\sum_{i=1}^{r_m}\|Q_i^{(m)}\|$.
\end{itemize}
In summary, we always have $\|P^{(k)}\|\leq
\sum_{m=k+1}^n\sum_{i=1}^{r_m}\|Q_i^{(m)}\|$ where
$r_{k+1}+\cdots+r_{n}\leq n-k+1 \leq n$. \quad$\Box$

The following result shows that even the size of the monomials
occurring in the algorithms is nicely bounded.

\begin{cor}
Let $M$ be the set of distinct monomials which are contained in some
polynomial occurring in Algorithm {\bf MFTriSet} and $H=\sum_{m\in
M} \|m\|$.
Then, $H\leq \sum_{P\in\PS} \cls(P) \|P\|+1$ where $\PS$ is the
input of the algorithm.
\end{cor}

\begin{Proof}
From the proof of Theorem \ref{th-len1}, a polynomial $P$ occurring
in the Algorithm {\bf MFTriSet} must have form \bref{eq-o1} or
\bref{eq-o2}. Then, a monomials $m$ of $P$ must be either 1 or
contained in some $LQ^{(k)}$, where $Q^{(k)}$ is an input polynomial
with class $k$ and $L\in \OS_{k-i,k}$. Thus, $H$ is not bigger than
the sum of the length of all such $LQ$ and $1$. From Lemma
\ref{lm-o2}, $\sum_{L_{i_2}\in
\OS_{2,k}}\|L_{i_2}Q^{(k)}\|+\cdots+\sum_{L_{i_k}
\in\OS_{k,k}}\|L_{i_k}Q^{(k)}\|+\|Q^{(k)}\|\leq k\|Q^{(k)}\|$.
Considering  all input polynomials $P$ and $1$, we get the
corollary.
\end{Proof}

\subsection{Complexity Analysis of {\bf MFTriSet}}

For a polynomial set $\PS$, we define $\tdeg(\PS)$ to be the highest
total degree of the elements in $\PS$. In this section, we will
always consider a Boolean polynomial set $\PS$ with $l$ polynomials
and $\tdeg(\PS)=d$.

\begin{theorem}\label{th-mfc2}
For an input polynomial set $\PS$ with $|\PS|=l$ and $\tdeg(\PS)=d$,
the bitsize complexity of {\bf MFTriSet} is $O(ln^{d+1}\sum_{P\in
\PS}\term(P))$. If $l\geq n$, the bitsize complexity of {\bf
MFTriSet} is $O(ln^{d+2}M)$ where $M=\max_{P\in\PS}{\term(P)}$.
\end{theorem}

As a consequence, Algorithm {\bf MFTriSet} is a polynomial-time
algorithm for a small $d$. For all the examples in Section 6, we
have $d\le 4$ and $n$ ranges from $40$ to $128$.
For such examples, the complexity is $O(n^8M)$ since $l$ is roughly
$O(n^2)$.

We will prove Theorem \ref{th-mfc2} in the rest of this section.
As in Section \ref{sec-mf2}, we assume that in the $k$-th round  of
{\bf MFTriSet} started as step 2, we deal with the polynomials of
class $k$, which is the worst case. Suppose that we have $l_k$
polynomials with class $k$ in the $k$-th round. Since the complexity
of computing $I+1$ is smaller than that of doing the polynomial
additions, we only consider the addition of two polynomials. Then we
need to do $l_k-1$ polynomial additions in order to eliminate $x_k$.
Thus, if we can estimate the number of the polynomials in $\PS$ in
every round, then we can obtain the complexity bound of {\bf
MFTriSet}. Note that, in Step 2.5 of {\bf MFTriSet}, we choose a $Q$
with the lowest degree, which is important for the complexity
analysis.

Suppose that we have a polynomial set $\SS=\{P_1,\ldots,P_l\}$ with
class $n$, which is the worst case. After eliminating $x_n$, we
obtain two sets of polynomials:
 $$\SS_{J}=\{\OJ_n P| P\in \SS \},\SS_{R}=\{\OR_n(P_s+P)| P \in \SS\}$$
where $P_s$ is a fixed polynomial with lowest degree in $\SS$ and
$\{\OJ_n$,$\OR_n\}$ are the operators defined in \bref{eq-rj}.
Note that $\tdeg(\SS_J)\leq d-1$ and $\tdeg(\SS_R)\leq d$. Moreover,
$|\SS_J|\leq l$ and $|\SS_R|\leq l$.
After eliminating $x_{n-1}$, we have four polynomial sets:
 \begin{eqnarray*}
 && \SS_{JJ}=\{\OJ_{n-1} P| P\in \SS_J\} , \SS_{JR}=\{\OJ_{n-1} P| P\in
 \SS_R\},\\
 && \SS_{RJ}=\{\OR_{n-1} (P_s+P)| P\in \SS_J \},\SS_{RR}=\{\OR_{n-1} (P_s+P)| P\in \SS_R \}.
 \end{eqnarray*}
Similarly, $|\SS_{JJ}|,|\SS_{RJ}|\leq |\SS_J| \leq l $ and
$|\SS_{JR}|,|\SS_{RR}|\leq |\SS_R| \leq l $. Since $P_s$ is a
polynomial with the lowest degree, we have $\tdeg(\OR_{n-1}
(P_s+P))\leq \tdeg(P)$ which means that $\tdeg(\SS_{RR})\leq
\tdeg(\SS_R)$ and $\tdeg(\SS_{RJ})\leq \tdeg(\SS_J)$. For the other two
sets, we can conclude $\tdeg(\SS_{JJ})\leq \tdeg(\SS_{J})-1 \leq d-2$
and $\tdeg(\SS_{JR})\leq \tdeg(\SS_R)-1 \leq d-1$.

Recursively, we have the following sequence
\begin{equation}\label{sq-s}
(\SS)\rightarrow (\SS_J,\SS_R) \rightarrow
(\SS_{JJ},\SS_{JR},\SS_{RR},\SS_{RJ})\rightarrow \cdots
\end{equation}
 For a set
$\SS_{O_1O_2\cdots O_k}$ where $O_i$ is $J$ or $R$, we have
$|\SS_{O_1O_2\cdots O_k}|\leq l$. We can deduce that
$\tdeg(\SS_{O_1O_2\cdots O_k})\leq d-s$ where $s$ is the number of
$O_i$ which is $J$. Therefore, the number of $J$ occurring in the
subscript of $\SS$ can be $d-1$ at most. As a consequence, in round
$n-k$ corresponding to the $(k+1)$-th part of the sequence
\bref{sq-s}, the number of $\SS_i$ is at most
$(^k_0)+(^k_1)+\cdots+(^{\ k}_{d-1})$. Thus, the number of
polynomials in round $n-k$ is at most $l(\sum_{i=0}^{d-1}(^k_i))$.
It implies that we need at most
$l(\sum_{k=0}^{n-1}\sum_{i=0}^{d-1}(^k_i))=l(\sum_{i=1}^{d}(^n_i))$
polynomial additions in the algorithm. It is easy to prove that in
other simpler cases,  the times of additions are still bounded by
$l(\sum_{i=1}^{d}(^n_i))$ or $O(ln^d)$.

Now let us estimate the complexity of polynomial additions in {\bf
MFTriSet}. We can define an operator $\OI_k$ as follows: If
$\cls(P)=k$, $\OI_k(P)=\initial(P)$; if $\cls(P)<k$, $\OI_k(P)=0$.
It is easy to prove that if we substitute $\OJ_i$ with $\OI_i$ in
equation \bref{eq-o1} and equation \bref{eq-o2} of Section 5.2, any
of the two equations will either be unchanged or become itself plus
one. Now we use $\term(P)$ to denote the number of monomials
occurring in $P$. Then we have $\term(\OI P)+\term(\OR P)\leq
\term(P)$. Similar to the proof of Theorem \ref{th-len1}, we can
prove the following lemma
\begin{lemma}
Let $n$ be the number of variables and $\PS$ the input of Algorithm
{\bf MFTriSet}.
Then, for any polynomial $T$ occurring in {\bf MFTriSet}, we have
$\term(T)\leq \sum_{P\in\PS}\term(P)+1$.
If $|\PS|>n$, then there exist $n$ polynomials $P_1,\ldots, P_n$ in
$\PS$ such that $\term(T)\leq
\term(P_1)+\term(P_2)+\cdots+\term(P_n)+1$.
\end{lemma}

Note that the bitsize complexity of computing the sum of $P_1$ and
$P_2$ is $O(n(\term(P_1)+ \term(P_2)))$.
Then the complexity of Algorithm {\bf MFTriSet} is $O(ln^{d+1}
(\sum_{P\in\PS}\term(P)))$. We have proved Theorem \ref{th-mfc2}.

\section{Experimental Results}
\label{sec-exp}

We have implemented algorithms {\bf TDCS} and {\bf MFCS} in $\R_2$
with the C language and tested them with a large number of
polynomial systems. In order to save storage space, we use the SZDD
to store the polynomials in our implementation \cite{zdd}.

For comparison, we also use the Gr\"obner basis algorithm (F4) in
Magma with Degree Reverse Lexicographic order, denoted by {\bf GB},
to solve these polynomial systems. The experiments are done on a PC
with a 3.19GHz CPU, 2G memory, and a Linux OS. The running times in
the tables are all given in seconds.

\subsection{Boolean Matrix Multiplication Problem}

For two $n\times n$ Boolean matrices $A$ and $B$, if $AB=I$, by the
linear algebra we can deduce that $BA=I$, where $I$ is the $n\times
n$ identity matrix. However, if we want to check the conclusion by
reasoning, it will become an extremely difficult problem. This
challenge problem was proposed by Stephen Cook in his invited talk
at SAT 2004 \cite{sat-cook,book-cook}. The best known result was
that the problem of $n=5$ can be solved by SAT-solvers in about
800-2000 seconds. The problem of $n=6$ were still unsolved
\cite{aca-biere}.

Now we test our software for this problem by converting the problem
into the solving of a Boolean polynomial system.
By setting the entries of $A$ and $B$ to be $2n^2$ distinct
variables, we can obtain $n^2$ quadratic polynomials from $AB=I$.
Then we compute the Gr\"obner basis or the zero decomposition of
this polynomials, and check wether the polynomials generated by
$BA=I$ can be reduced to $0$ by the Gr\"obner basis or by every
characteristic set in the zero decomposition. In this way, we can
prove the conclusion.

We use the CS method to illustrate the above procedure. Let $\PS_1$
and $\PS_2$  be the polynomial sets generated by $AB=I$ and $BA=I$
respectively. With the CS method, we have
$$\hzero(\PS_1) = \cup_i \hzero(\A_i)$$
where $\A_i$ are triangular sets. If $\prem(P,\A_i)=0$ for all
possible $i$ and $P\in \PS_2$, then we have solved the problem. It
is clear that the major difficulty here is to compute the
decomposition.

For $n=4,5,6$, the numbers of variables are $32,50,72$ respectively.
Therefore, computing the Gr\"obner basis or the zero decomposition
of this polynomials will be a hard work. We used {\bf GB} and our
{\bf MFCS} algorithm to solve the problem with $n=4,5,6$. The
running time given in Table \ref{tab-00} includes solving the
equations generated by $AB=I$ and checking the conclusion $BA=I$.
Notation $\bullet$ means memory overflow.


\begin{table}[!ht]\centering
\begin{tabular}{|c|c|c|c|}\hline
    &n=4&n=5&n=6\\\hline
{\bf MFCS}& $0.11$  & $41$ & $196440$\\ \hline
{\bf GB} & $2363$  & $\bullet$  & $\bullet$\\ \hline
\end{tabular}
\caption{Running times for Boolean matrix multiplication problems}
\label{tab-00}
\end{table}

\subsection{Equations from Stream Ciphers Based on Nonlinear Filter Generators}

In this section we generate our equations from stream ciphers based
on LFSRs. We first show how these polynomial systems are generated.
A linear feedback shift register (LFSR) of length $L$ can be simply
considered as a sequence of $L$ numbers $(c_1, c_2,\ldots,c_{L})$
from $\F_2$ such that $c_L\ne0$ \cite{hac}. For an {\bf  initial
state} $S_0=(s_0,s_1,\ldots,s_{L-1})\in\F_2^{L}$, we can use the
given LFSR to produce an infinite sequence satisfying
\begin{equation}\label{eq-se} s_i = c_1
s_{i-1} + c_2 s_{i-2} + \cdots +c_{L} s_{i-L},
i=L,L+1,\cdots.\end{equation}
A key property of an LFSR is that if the related {\bf  feedback
polynomial} $P(x) = c_Lx^L + c_{L-1} x^{L-1}+\cdots+c_1x-1$ is
primitive, then the sequence \bref{eq-se} has period $2^L -1$
\cite{hac}. The number of non-zero coefficients in $P$ is called the
{\bf  weight} of $P$, denoted by $w_P$.

An often used technique in stream ciphers to enhance the security of
an LFSR is to add a {\bf  nonlinear filter} to the LFSR. Let
$f(x_1,\ldots,x_m)$ be a Boolean polynomial with $m$ variables. We
assume that $m \le L$. Then we can use $f$ and the sequence
\bref{eq-se} to generate a new sequence as follows
\begin{equation}\label{eq-ze}
z_t = f(s_{t+k_1},s_{t+k_2}\ldots,s_{t+k_m}),
  t=0,1,\ldots
\end{equation}
where $\{k_i\}_{1 \leq i \leq m}$ is called the {\bf  tapping
sequence}. A combination of an LFSR and a nonlinear polynomial $f$
is called a {\bf  nonlinear filter generator} (NFG).

The filter functions used in this paper are due to Canteaut  and
Filiol \cite{canfil1}:
{\small\begin{itemize} \item CanFil 1,\
$x_1x_2x_3+x_1x_4+x_2x_5+x_3$ \item CanFil 2,\
$x_1x_2x_3+x_1x_2x_4+x_1x_2x_5+x_1x_4+x_2x_5+x_3+x_4+x_5$ \item
CanFil 3,\ $x_2x_3x_4x_5+x_1x_2x_3+x_2x_4+x_3x_5+x_4+x_5$ \item
CanFil 4,\ $x_1x_2x_3+x_1x_4x_5+x_2x_3+x_1$ \item CanFil 5,\
$x_2x_3x_4x_5+x_2x_3+x_1$ \item CanFil 6,\
$x_1x_2x_3x_5+x_2x_3+x_4$ \item CanFil 7,\
$x_1x_2x_3+x_2x_3x_4+x_2x_3x_5+x_1+x_2+x_3$ \item CanFil 8,\
$x_1x_2x_3+x_2x_3x_6+x_1x_2+x_3x_4+x_5x_6+x_4+x_5$ \item CanFil
9,\ $x_2x_4x_5x_7+x_2x_5x_6x_7+x_3x_4x_6x_7+
 x_1x_2x_4x_7+x_1x_3x_4x_7+x_1x_3x_6x_7+x_1x_4x_5x_7+x_1x_2x_5x_7+
 x_1x_2x_6x_7+x_1x_4x_6x_7+x_3x_4x_5x_7+x_2x_4x_6x_7+x_3x_5x_6x_7+
 x_1x_3x_5x_7+x_1x_2x_3x_7+
 x_3x_4x_5+x_3x_4x_7+x_3x_6x_7+x_5x_6x_7+x_2x_6x_7+x_1x_4x_6+x_1x_5x_7+
 x_2x_4x_5+x_2x_3x_7+x_1x_2x_7+x_1x_4x_5+
 x_6x_7+x_4x_6+x_4x_7+x_5x_7+x_2x_5+x_3x_4+x_3x_5+x_1x_4+x_2x_7+
 x_6+x_5+x_2+x_1$
\item CanFil 10,\
$x_1x_2x_3+x_2x_3x_4+x_2x_3x_5+x_6x_7+x_3+x_2+x_1$.
\end{itemize}}

In the experiments, we use our algorithms to find
$S_0=(s_0,s_1,\ldots,s_{L-1})$ by solving the following equations
for given $c_i$, $z_i$, and $f$
\begin{equation}\label{eq-seq}
z_t = f(s_{t+k_1},s_{t+k_2}\ldots,s_{t+k_m}),
  t=0,1,\ldots,k\end{equation} where $k$ is a positive integer,
$s_i$ satisfy \bref{eq-se}, and $\{k_1,\ldots,k_m\}$ is a tapping
sequence.

We compare four different algorithms for solving these equations.
Two of them are the {\bf MFCS} and {\bf GB}.
Faug\`ere and Perret suggested to us that an incremental version of
the Gr\"obner basis algorithm is faster than {\bf GB} for the
equations generated by the LFSR.
%
Therefore, we also compare the incremental Gr\"obner basis algorithm
and the incremental {\bf TDCS}, denoted {\bf IGB} and {\bf ITDCS}
respectively.
%
%
Note that the F5 method \cite{f5} and the CS method presented in
\cite{maza1} also use the incremental technique.

Let $HS$ be the field polynomials $\{x_1^2+x_1,\ldots,x_n^2+x_n\}$
and $PS=\{P_1,P_2,\ldots,P_k\}$ be the input polynomials with $P_i$
be the polynomial generated from the i-th output bit. Then we
compute the {\bf IGB} by the following codes in Magma:

{\em R$<$x$_1,\ldots,$x$_n>$:=PolynomialRing(GF($2$),n,``grevlex");

HS:=[R.i$^{\wedge}2$+R.i: i in [1..Rank(R)]]; G:=HS;

for i:=1 to k do

G:=G cat [PS.i]; G:= GroebnerBasis(G);

end for;

G;
}

We did three sets of experiments with increasing difficulties.
The test problems are similar to those in \cite{gaof2} but are more
difficult. We also compare our method with one of the benchmark
implementations of the Gr\"obner basis method on the same computer,
which are not given in \cite{gaof2}.

In the first set of experiments, we choose a simple tapping sequence
$\{0,1,2,3,4,5,6\}$ and the feedback polynomials for $n=40, 60, 81,
100,128$ are respectively $x^{40}+x^{21}+x^{19}+x^2+1$,
$x^{60}+x^1+1$, $x^{81}+x^4+1$, $x^{100}+x^{37}+1$,
$x^{128}+x^{29}+x^{27}+x^2+1$. The results are given in Table
\ref{tab-1}, where $L$ is the number of variables, $k$ is the number
of equations (see \bref{eq-seq}). $k$ is the smallest number such
that the system has a unique solution, $w_P$ is the weight of the
feedback polynomial $P$, and $\bullet$ means memory overflow.

{\footnotesize
\begin{table}[!ht]\centering
\begin{tabular}{|c|c|c|c|c|c|c|}\hline
Filters & L($w_f$)=& 40 ($5$)& 60 ($3$) & 81 ($3$)& 100 ($3)$& 128 ($5$) \\
\hline%
   &{\bf MFCS}& 0.10& 0.02& 0.07 & 0.37& 0.49\\
&{\bf ITDCS}& 0.10& 0.04& 0.05& 0.21&0.37 \\
CanFil1&{\bf IGB}&   0.42&0.99 & 2.29& 3.26& 8.32\\
   &{\bf GB}  & 0.91& 0.43& 8.12 & 3.61& 1997.2\\
\cline{2-7}  &k &52& 114&154 & 140& 230 \\\hline
   & {\bf MFCS}& 0.17& 0.03& 0.07& 0.59& 1.11\\
   & {\bf ITDCS}& 0.04& 0.02& 0.06& 0.19& 0.53\\
CanFil2 & {\bf IGB}& 0.43& 0.65& 1.61& 3.17& 7.13\\
   &{\bf GB}  & 0.92& 30.65& 0.02& 55.09& $\bullet$\\
\cline{2-7}&k &44 &72& 138 & 140& 217 \\\hline
  &{\bf MFCS}& 0.17& 0.03& 0.07 & 0.59& 1.11\\
&{\bf ITDCS}& 0.14& 0.03& 0.23& 1.10&0.72 \\
CanFil3&{\bf IGB}& 0.16&0.96 & 2.51& 6.04& 16.08\\
 & {\bf GB} &178.57& 1.68& $\bullet$& $\bullet$ & $\bullet$\\
\cline{2-7}&k &64& 114&162& 120& 128 \\\hline
 &{\bf  MFCS}& 0.09& 0.05& 0.07 & 0.83& 2.70\\
&{\bf ITDCS}& 0.14& 0.09& 0.09& 2.91&2.01 \\
CanFil4&{\bf IGB}&   0.17&0.89 & 1.99& 2.13& 10.26\\
 & {\bf GB   }&0.65 & 2.24& 0.39&$\bullet$&$\bullet$\\
\cline{2-7}&k &60& 168& 154 &150& 180  \\\hline
&{\bf MFCS}& 0.03& 0.01& 0.03 & 0.08& 0.12\\
&{\bf ITDCS}& 0.04& 0.05& 0.11& 0.18&0.59 \\
CanFil5&{\bf IGB}&   0.14&0.37 & 0.80& 1.59& 3.46\\
& {\bf GB   } &0.10  & 0.06 & 0.10&0.50& 0.85\\
\cline{2-7}&k &40& 60 & 81 & 100& 128 \\
\hline
&{\bf MFCS}& 0.05& 0.04& 0.08 & 0.11& 0.35\\
&{\bf ITDCS}& 0.09& 0.04& 0.10& 0.29&1.07 \\
CanFil6&{\bf IGB}&   0.08&0.35 & 0.80& 1.70& 5.28\\
& {\bf GB }& 0.24& 0.09& 0.01& 0.65& $\bullet$\\
\cline{2-7}&k &52 & 108&146 & 160& 230 \\\hline
&{\bf  MFCS}& 0.05& 0.02& 0.08 & 0.38& 0.70\\
&{\bf ITDCS } & 0.03& 0.03& 0.08& 0.24&0.42 \\
CanFil7&{\bf IGB}&   0.10&0.81 & 1.86& 3.32& 9.78\\
& {\bf GB   }&0.27  & 0.40&0.01& 831.89& $\bullet$\\
\cline{2-7}&k & 40 & 120 & 154 & 150 & 218\\\hline
&{\bf MFCS}& 0.32& 0.08& 0.21 & 0.61& 1.31\\
&{\bf ITDCS}& 0.09& 0.06& 0.14& 0.25&0.66 \\
CanFil8&{\bf IGB}&   0.13&0.30 & 1.26& 2.09& 6.11\\
&{\bf GB    } &0.88 & 0.56 & 92.51&20.03& $\bullet$\\
\cline{2-7}&k &44& 60& 154 & 140 & 218\\\hline
&{\bf  MFCS}& 2.94& 0.30& 0.64 & 0.79& 15.31\\
&{\bf ITDCS}& 0.45& 0.06& 0.24& 1.22&1.28 \\
CanFil9&{\bf IGB}&   4.39&5.13 & 13.15& 17.78& 47.62\\
&{\bf GB   } &$\bullet$ & 90.49 &$\bullet$& $\bullet$& $\bullet$\\
\cline{2-7}&k &48& 102 & 113 & 110  & 218\\\hline
&{\bf  MFCS}& 0.39& 0.06& 0.12 & 1.40& 3.43\\
&{\bf ITDCS}& 0.12& 0.04& 0.12& 0.57&0.49 \\
CanFil10&{\bf IGB}&   4.48&28.16 & 50.87& 63.63& 100.39\\
&{\bf  GB }  &28.72  & 2.21 & 492.16& $\bullet$ & $\bullet$ \\ 
\cline{2-7}&k & 44&90 & 122 & 140 & 205\\\hline
\end{tabular}
\caption{Examples with simple feedback polynomials and tapping
sequences} \label{tab-1}
\end{table}
}

In the second set of experiments,  we generate more difficult
equations in the cases of $L=40$ and $k=60$ by changing the feedback
polynomial to
$x^{40}+x^{35}+x^{32}+x^{27}+x^{24}+x^{19}+x^{15}+x^{12}+x^{7}+x^1+1$.
The results are given  in Table \ref{tab-3}.

\begin{table}[!ht]\centering
\begin{tabular}{|c|c|c|c|c|}\hline
Filter&{\bf ITDCS}&{\bf MFCS}&{\bf IGB}&{\bf GB}\\\hline
Canfil1&0.78&2.44&0.89&55.73 \\ \hline
Canfil2&0.47&2.17&0.66&49.33 \\ \hline
Canfil3&1.01&8.10&3.16&$\bullet$ \\ \hline
Canfil4&0.99&2.24&0.62& 26.10\\ \hline
Canfil5&0.58&2.80&3.00& $\bullet$\\ \hline
Canfil6&0.58&2.14&2.81&$\bullet$ \\ \hline
Canfil7&0.16&0.35&0.27& 16.64\\ \hline
Canfil8&0.26&5.81&0.34&33.35\\ \hline
Canfil9&6.83&75.62&8.54& $\bullet$\\ \hline
Canfil10&0.70&3.04&4.87&$\bullet$\\ \hline
\end{tabular}
\caption{Examples with larger feedback polynomials} \label{tab-3}
\end{table}

In the third set of experiments,  we generate more dense polynomial
systems by changing the tapping sequence.  The results are given in
Table \ref{tab-4}, in which $L=40$, $k=55$, the feedback polynomial
is $x^{40}+x^{37}+x^{34}+x^{21}+x^{11}+x^5+1$ and the tapping
sequence is $\{0,6,11,18,25,31,37\}$. And $*$ means that we have computed
over 2 hours and did not obtain the solutions.

\begin{table}[!ht]\centering
\begin{tabular}{|c|c|c|l|}\hline
Filter&{\bf MFCS }&{\bf ITDCS}&{\bf IGB}\\\hline
Canfil1&109.91&*& $\bullet$ after 10m\\ \hline
Canfil2&160.98&*&$\bullet$ after 8m\\ \hline
Canfil3&149.05&*&$\bullet$ after 28m \\ \hline
Canfil4&11.19&*&$\bullet$ after 60m\\ \hline
Canfil5&23.98&*&$\bullet$ after 4m\\ \hline
Canfil6&107.39&*&$\bullet$ after 6m \\ \hline
Canfil7&13.95&*&$\bullet$ after 37m\\ \hline
Canfil8&855.04&*&$\bullet$ after 60m\\ \hline
\end{tabular}
\caption{Examples with larger feedback polynomials and nontrivial
tapping sequences} \label{tab-4}
\end{table}

From the experiments, we have the following observations.
\begin{itemize}
\item
From Table \ref{tab-1},  we can see that for these ``simple"
examples, {\bf ITDCS} is the fastest method.
{\bf IGB} and {\bf MFCS} are also very efficient with {\bf MFCS}
better than {\bf IGB} in most cases. {\bf GB} tends to generate
large polynomials and causes memory overflow.

\item
From Table \ref{tab-3}, we can see that for these ``moderately difficult"
polynomial systems, {\bf ITDCS} is still the fastest method. Now,
{\bf IGB} performs better than {\bf MFCS}.

\item
From Table \ref{tab-4}, we can see that for the ``most difficult"
polynomial systems, {\bf MFCS} is the only algorithm that can find
the solutions on our computer. {\bf IGB} and {\bf GB} quickly use
all the memory and cause memory overflow. {\bf ITDCS} has been run
for two hours without giving a result.
The reason is that, in this case,  {\bf ITDCS} and {\bf IGB} need to
deal with some high degree and dense polynomials. On the other hand,
due to  Theorems \ref{th-len1} and \ref{th-mfc2}, the polynomials
occurring in Algorithm {\bf MFCS} are much smaller.

\end{itemize}

In summary, Algorithm {\bf MFCS} seems to be the most efficient and
stable approach to deal with these kinds of polynomial systems. The
main reason is that the size of the polynomials in this algorithm is
effectively controlled due to Theorems \ref{th-len1} and
\ref{th-mfc2}.
To use SZDD \cite{zdd} to represent polynomials is another key
factor in memory saving. Note that SZDD suits the CS method very
well. The CS method will generate a large number of components and
the polynomial sets representing different components differ only
for a very few number of polynomials due to the way of generating
new components (see Step 2.6.3 of Algorithm 4.3). Then different
polynomial sets will share memory for their common polynomials, and
as a consequence, the total memory consumption is well contained.

\begin{table}[!ht]\centering
\begin{tabular}{|c|c|c|c|c|c|c|c|c|}\hline
&Canfil1&Canfil2&Canfil3&Canfil4&Canfil5&Canfil6&Canfil7&Canfil8\\\hline
$N_C$&13749&23881&7251&1657&1086&3331&1551&180710 \\
\hline
$R\approx$ & $2^{-26}$& $2^{-25}$& $2^{-27}$ & $2^{-29}$&$2^{-30}$& $2^{-28}$& $2^{-29}$& $2^{-24}$\\
\hline
\end{tabular}
\caption{The number of components for the examples in Table
\ref{tab-4}} \label{tab-5}
\end{table}

For Algorithm {\bf MFCS}, the bottle neck problem is how to control
the number of components (that is, the number of polynomial sets in
$\PS^*$ in the output of Algorithm {\bf MFTriSet}). Theoretically,
this number is exponential in the worst case. Practically, this
number could also be very large. But, comparing to the number $2^n$
of exhaust search, the number of components generated in {\bf
MFTriSet} is still very small.
In Table \ref{tab-5}, we give the numbers of components for each
example in Table \ref{tab-4}. In this table, $N_C$ is the number of
components and $R=\frac{N_C}{2^n}$ could be considered as a measure
of effectiveness of Algorithm {\bf MFTriSet}. We can see that $R$ is
very small for all examples.

\subsection{Attack on Bivium-A}
Bivium is a simple version of the eStream stream cipher candidate
Trivium \cite{estream} . It is built on the same design principles
of Trivium. The intention is to reduce the complexity of Trivum, and
to extend the attacks on Bivium to Trivium. Bivium has two versions
Bivium-A and Bivium-B. Here we focus on attacking Bivium-A. There
have been several successful attacks on Bivium-A, and we want to
show that our algorithm is comparable with these algorithms.

The Bivium-A is given by the following pseudo-code:

\begin{center}
for $i=1$ to $N$ do

\begin{tabular}{rcl}
 $t_1$ & $\leftarrow$& $s_{66}+s_{93}$\\
 $t_2$ &$\leftarrow$& $s_{162}+s_{177}$\\
 $z_i$ &$\leftarrow$& $t_2$\\
 $t_1$ &$\leftarrow$& $t_1+s_{91}\cdot s_{92}+s_{171}$\\
 $t_2$ &$\leftarrow$& $t_2+s_{175}\cdot s_{176}+s_{69}$\\
 $(s_1,s_2,\ldots,s_{93})$ &$\leftarrow$& $(t_2,s_1,\ldots,s_{92})$\\
$(s_{94},s_{95},\ldots,s_{177})$& $\leftarrow$&
$(t_1,s_{94},\ldots,s_{176})$
\end{tabular}
\end{center}
We want to recover the initial state $(s_1,\ldots,s_{177})$ from the
given $N$ output bits $(z_1,\ldots,z_{N})$. Note that the degree of
the equations will increase after several clocks. In order to avoid
this problem,  we can introduce two new variables and two equations
for each clock:
\begin{eqnarray}
 s_{178}= && s_{66}+s_{93}+s_{91}\cdot s_{92}+s_{171}\\
 s_{179}= && s_{162}+s_{177}+s_{175}\cdot s_{176}+s_{69}
\end{eqnarray}
Then we can obtain a Boolean polynomial system with $2N+177$
variables and $3N$ equations.

The results of the successful attacks on Bivium-A
\cite{bivium-mc,bivium-radd,bivium-fg}\footnote{In \cite{bivium-fg},
they give four different results by solving in different ways. Here
we only list the result by adding new variables but without guessing
any variables.} is given in Table \ref{tab-bv1}.

\begin{table}[!ht]\centering
\begin{tabular}{|c|c|c|c|}\hline
Method& Graph for sparse system & SatSolver & Gr\"obner Basis

\\ \hline
Time&``about a day"& 21 sec&400 sec \\
\hline Output Bits& 177& 177 & 2000\\ \hline
\end{tabular}
\caption{The known results for Bivium-A} \label{tab-bv1}
\end{table}

In our experiments, we use the algorithm {\bf MFCS} and the
equations are generated by adding two new variables for each clock.
We run {\bf MFCS}  on a sample of 100 different random initial
states. We observed that the different initial keys make a great
difference to the results. For every initial state, we can find a
number $M$. When the number of output bits $N$ is not less than $M$,
the equations can be solved within one minute. When $N$ becomes much
bigger, the running time will increase slowly. However, if $N$ is
less than $M$, the running time will be much longer than one minute.
From our experiment results, the value of $M$ is from $200$ to
$700$. In our experiments, we set $N=700$.
%

The average time for solving the problem by {\bf MFCS} with $700$
output bits is 49.3 seconds. We also tried to use {\bf GB} to solve
the same sample by the same computer. The equations are also
generated by adding two variables for each clock. In order to solve
the equations, we need $1700$ output bits. If the output is less
than $1700$ bits, the memory will be exhausted. For $N=1700$, the
average time for solving the problem by {\bf GB} is 303.3 seconds.
If we set $N=2000$ as in \cite{bivium-fg}, the average time is 521.6
seconds. From the results, we can see that our algorithm is
comparable with the known successful algorithms in this problem.

\section{Conclusions}

In this paper, we present two algorithms {\bf TDCS} and {\bf MFCS} to solve nonlinear equation
systems in finite fields based on the idea of characteristic set.
Due to the special property of finite fields, the given algorithms
have better properties than the general characteristic set method.
In particular, we obtain an explicit formula for the number of
solutions of an equation system, and give the bitsize complexity of
Algorithm {\bf TDCS} for Boolean polynomials. We also prove that the size
of the polynomials in {\bf MFCS} can be effectively controlled,
which allows us to avoid the expression swell problem effectively.

We test our methods by solving polynomial systems generated by the
Boolean matrix problem, stream cipher Bivium-A and stream ciphers
based on nonlinear filter generators. All these equations have block
triangular structure. Extensive experiments show that our methods
are efficient for solving this kind of equations and Algorithm {\bf
MFCS} seems to be the most efficient and stable approach for these
problems.

The experiments are only done for Boolean polynomials in this paper.
It our future work to see whether the algorithms proposed in this paper
can be developed into practically efficient software packages for finite fields
other than $\F_2$.
It is expected that elimination techniques developed in previous work
on CS methods will also be needed.

{\vskip10pt\noindent\bf Acknowledgements}. We want thank the anonymous referees for helpful comments and suggestions.

\end{document}